\documentclass[a4paper]{aa}
\usepackage{txfonts}
\usepackage[usenames,dvipsnames,svgnames,table]{xcolor}
\usepackage{graphicx}
\usepackage{hyperref}
\usepackage[hyphenbreaks]{breakurl}

\outer\def\gtae {$\buildrel {\lower3pt\hbox{$>$}} \over 
{\lower2pt\hbox{$\sim$}} $}
\outer\def\ltae {$\buildrel {\lower3pt\hbox{$<$}} \over 
{\lower2pt\hbox{$\sim$}} $}
\newcommand{\Msun} {M$_{\odot}$} 
\newcommand{\Rsun} {R$_{\odot}$}
\newcommand{\Swift}{{\sl Swift}}
\newcommand{\LISA}{{\sl LISA}}
\newcommand{\Gaia}{{\sl Gaia}}
\newcommand{\HST}{{\sl HST}}
\newcommand{\xmm}{{\sl XMM-Newton}}
\newcommand{\galex}{{\sl GALEX}}
\newcommand{\Porb}{P_{\rm orb}}

\newcommand{\tickYes}{\checkmark}
\newcommand{\tickNo}{\hspace{1pt}\ding{55}}

\usepackage{xspace}
\usepackage{amsfonts}
\usepackage{pifont}

\begin{document}
\title{The physical properties of AM CVn stars: new insights from Gaia DR2}
\author{G. Ramsay\inst{1}, M.\ J. Green\inst{2}, T.\ R. Marsh\inst{2}, T. Kupfer\inst{3,4,5}, E. Breedt\inst{6}, V. Korol\inst{7},  P.J. Groot\inst{8}, C. Knigge\inst{9},\\ G. Nelemans\inst{8,10}, D. Steeghs\inst{2}, P. Woudt\inst{11}, A. Aungwerojwit\inst{12}}

\authorrunning{Ramsay et al.}
\titlerunning{AM~CVn stars from \Gaia\, DR2}
\institute{Armagh Observatory and Planetarium, College Hill, Armagh, BT61 9DG, UK\label{inst1}\and
  Department of Physics, University of Warwick, Coventry CV4 7AL, UK\label{inst2}\and  
  Kavli Institute for Theoretical Physics, University of California, Santa Barbara, CA 93106, USA\label{inst3}\and
  Department of Physics, University of California, Santa Barbara, CA 93106, USA \label{inst4}\and
  Division of Physics, Mathematics and Astronomy, California Institute of Technology, Pasadena, CA 91125, USA\label{inst5}\and 
  Institute of Astronomy, University of Cambridge, Madingley Road, Cambridge CB3 0HA, UK\label{inst6}\and
  Leiden Observatory, Leiden University, PO Box 9513, 2300 RA, Leiden, the Netherlands\label{inst7}\and
  Department of Astrophysics/IMAPP, Radboud University, PO Box 9010, NL-6500 GL Nijmegen, Netherlands\label{inst8}\and
  Department of Physics and Astronomy, University of Southampton, Southampton SO17 1BJ, UK\label{inst9}\and
Institute of Astronomy, KU Leuven, Celestijnenlaan 200D, B-3001
Leuven, Belgium\label{inst10}\and
Inter-University Institute for Data Intensive Astronomy, Department of Astronomy, University of Cape Town, Private Bag X3,  Rondebosch 7701, South Africa\label{int11}\and
Department of Physics, Faculty of Science, Naresuan University, Phitsanulok 65000, Thailand\label{int12}\\
\email{gavin.ramsay@armagh.ac.uk}}

\date{Accepted: Oct 14 2018}

\abstract {AM~CVn binaries are hydrogen deficient compact binaries
  with an orbital period in the 5--65 min range and are predicted to
  be strong sources of persistent gravitational wave radiation. Using
  \Gaia\, Data Release 2, we present the parallaxes and proper motions
  of 41 out of the 56 known systems. Compared to the parallax
  determined using the \HST\, Fine Guidance Sensor we find that the
  archetype star, AM~CVn, is significantly closer than previously
  thought. This resolves the high luminosity and mass accretion rate
  which models had difficulty in explaining. Using Pan-STARRS1 data we
  determine the absolute magnitude of the AM~CVn stars. There is some
  evidence that donor stars have a higher mass and radius than
  expected for white dwarfs or that the donors are not white
  dwarfs. Using the distances to the known AM~CVn stars we find strong
  evidence that a large population of AM~CVn stars have still to be
  discovered. As this value sets the background to the gravitational
  wave signal of \LISA\, this is of wide interest. We determine the
  mass transfer rate for 15 AM~CVn stars and find that the majority
  have a rate significantly greater than expected from standard
  models. This is further evidence that the donor star has a greater
  size than expected.}

\keywords{
Physical data and processes: accretion, accretion disc: stars:
distances stars}

\maketitle
   
%----------------------------------------------------------------------
\section{Introduction}
%----------------------------------------------------------------------

AM~CVn stars occupy the extreme short period tail of stellar binaries,
with observed orbital periods in the range $\sim5-65$~min. They
consist of white dwarfs accreting material from Roche lobe-filling
companion stars, typically a lower mass white dwarf or a
semi-degenerate helium-rich star. The mass transfer in these binaries
is driven by gravitational wave radiation and they are expected to be
strong sources of low frequency gravitational waves
\citep{Nelemans2004}. In particular, the fact that their binary
properties such as orbital period and mass ratio can be measured from
electromagnetic observations makes them useful as `verification
sources' for the Laser Interferometer Space Antenna (\LISA)
mission. In a separate paper, we predict the gravitational wave strain
of the 16 currently-known \LISA\, verification binaries, 11 of which
are AM~CVn stars \citep{Kupfer2018}.

There are three proposed channels for the formation of AM~CVn stars,
but their relative importance is not yet clear. The population
synthesis models of \citet{Nelemans2001a} suggest that the majority of
AM~CVn stars form from double white dwarf binaries that evolve closer
together as a result of gravitational wave radiation, and start mass
transfer at orbital periods of $\sim$2--3~min
\citep{Paczynski67}. Given these small orbital separations, the
initial mass transfer occurs as direct impact, which will lead to
unstable mass transfer and hence merger in the majority of cases. The
fraction of double white dwarfs which survive to become stable mass
transferring AM~CVn stars is highly uncertain, and depends on the
efficiency with which the spin of the accretor can be tidally coupled
to the binary orbit to stabilise the mass transfer
\citep{Marsh2004}. There has also been a suggestion by
\citet{Shen2015} that due to increased friction from ejected material
in nova eruptions earlier in the evolution of the binary, all double
white dwarf binaries will merge, and that no AM~CVn stars should form
in this way.

The alternative is that the companion is a core helium-burning star
instead of another white dwarf
\citep{Savonije1986,IbenTutukov1987}. Such a binary will reach a
minimum period of $\sim10$~min before mass transfer starts and the
helium star will get increasingly degenerate as it evolves to longer
periods.

A third channel involves a hydrogen-rich cataclysmic variable with a
donor that is already partially evolved at the onset of mass
transfer. The binary then loses its hydrogen through evolution and
accretion to become an AM~CVn star with a period at the long end of
the range. It is considered an unimportant channel compared to the
white dwarf and helium star scenarios given the long evolutionary
timescales involved, but potential progenitor binaries for this
channel have been identified \citep{Breedt2012,Carter2013b}.

Only the two eclipsing AM~CVn stars, YZ~LMi (SDSS\,J0926+3624) and
Gaia14aae, have parameters measured to high enough precision to
discriminate between these models. YZ~LMi is likely to be of helium
star origin, although the white dwarf route cannot be ruled out
completely \citep{Copperwheat2011}. Gaia14aae on the other hand, is
inconsistent with a white dwarf donor scenario, but it is also not
straightforwardly compatible with either of the other models
\citep{Green2018_14aae}.  A detailed discussion of the three formation
channels and a review of the observed properties of the AM~CVn
population is given by \citet{Solheim2010}.

A well-defined sample of AM~CVn stars from the Sloan Digital Sky
Survey (SDSS) made it possible to compare the observed space density
of these binaries, $\rho = (5\pm3)\times10^{-7}$ pc$^{-3}$, to
population synthesis predictions \citep{Carter2013a}. Even though the
models take into account a range of efficiencies for the spin-orbit
coupling and the subsequent AM~CVn survival rate, the models
overpredict the space density by an order of magnitude. The reason for
the discrepancy is not clear, but may be related to the uncertainty of
the distribution of these binaries in the Galaxy
\citep{Nissanke2012}. Most surveys where AM~CVn stars have been found
have covered only high Galactic latitudes, and it is possible that a
substantial fraction reside in the Galactic plane.

One of the main limitations in modelling the spatial distribution of
the AM~CVn population and calibrating models of space density and
luminosity is the lack of accurate distances. Only five systems have a
parallax determined using the Hubble Space Telescope Fine Guidance
Sensor \citep[{\sl HST FGS};][]{Roelofs2007c} with distances to others
are generally estimated by comparing model fluxes with
observations. With reliable distances we can determine the mass
transfer rate, $\dot{M}$, which in conjunction with an orbital period,
can constrain which of the three formation channels the binary
formed. In turn, this can provide a more reliable value for their
space density.

The \Gaia\, Data Release 2 (DR2) on 25~April~2018 provided the
parallaxes of 1.3 billion stars down to $G\sim21$
\citep{GaiaBrown2018} and has allowed us to determine the distances to
41 of the 56 known AM~CVn stars. The first \Gaia\, Data Release
\citep[DR1;][]{GaiaPrusti2016} in September 2016 relied on a combined
      {\sl Tycho-Gaia}\, astrometric solution
      \citep[TGAS;][]{GaiaBrown2016}, and did not include any AM~CVn
      stars. However, it included parallaxes of 16 hydrogen
      cataclysmic variables \citep[CVs;][]{Ramsay2017}, which provided
      a validation of the Disk Instability Model which is widely used
      to model accreting binaries, including AM~CVn stars
      \citep{Osaki1989,Kotko2012,CannizzoNelemans2015}. Another key
      result from this work was the comparison between the {\sl HST
        FGS}, Very Large Array radio data and \Gaia\, DR1 parallax
      measurements of the CV SS\,Cyg, which resolved a long-standing
      discrepancy in the distance to (and hence luminosity of) this
      system, and showed that the \HST\, parallaxes may be unreliable.

In this paper, we use the GDR2 parallax measurements to determine the
absolute magnitudes of AM~CVn stars and their mass accretion rates. We
then use these results to infer the space density of these binaries.

\section{The known AM~CVn stars}

Observationally, AM~CVn stars are characterised by their hydrogen
deficient optical spectra and blue colour, so surveys for AM~CVn stars
have typically focussed on these properties to identify new members of
the class. The past decade has seen a rapid increase in the number of
known AM~CVn stars. Firstly because of a dedicated spectroscopic
survey of colour-selected targets from SDSS \citep[][and references
  therein]{Carter2013a} and secondly from photometric and
spectroscopic follow-up of transient events in large area photometric
surveys, such as the Catalina Real-time Transient Survey
\citep[CRTS;][]{Breedt2014}, the Palomar Transient Factory
\citep[PTF;][]{Levitan2015} and the All-Sky Automated Survey for
Supernovae \citep[ASASSN;][]{Breedt2015}.

Since the data compilation by \citet{Levitan2015}, a number of
additional AM~CVn systems have been discovered. Some of the more
recent discoveries include Gaia14aae, the first in which the white
dwarf is fully eclipsed \citep{Campbell2015,Green2018_14aae},
ASASSN-15fp, the longest period AM~CVn system to have been observed in
outburst so far \citep{Cartier2017ePESSTO,Marsh2017}, and
SDSS\,J1351-0643, the first system with a period shorter than 17~min
to be discovered in $\sim$15~years \citep{Green2018a}. In
Table~\ref{sources}, we list the 56 AM~CVn stars known at present,
ordered by increasing orbital period. We provide a full table,
including J2000.0 and J2015.5 sky coordinates and multi-wavelength
photometry in the online material. The column description of the full
table is shown in the appendix.

Time series spectroscopy remains the most reliable method to measure
the orbital periods of AM~CVn stars, but it is a challenging task due
to the faintness of many systems and the short exposures which are
needed to phase resolve the short orbital period. For systems which
display outbursts (see Table~\ref{sources}) the superhump period may
be used as a proxy. These are flux variations observed during
superoutbursts, resulting from the interaction between the precessing
accretion disc and the donor star \citep[e.g.][]{Wood2011}. It is
typically a few per cent longer than the orbital period.  Other
proxies include the relationship between the equivalent width of
emission lines and the orbital period \citep{Carter2013a}, and the
recurrence time between outbursts \citep{Levitan2015}.

The orbital periods of the known AM~CVn stars range from 5.4~min to
65.6~min. Eight of the known systems do not have an estimate of their
orbital period yet. AM~CVn stars at the short period end
($\Porb<20$\,min), are akin to novalike CVs, with hot, high state
accretion discs. The accretion rate drops as the binary evolves to
longer periods, and at the longest periods in the range (i.e. lowest
accretion rates) the discs are in a low, stable state. At intermediate
periods, $20<\Porb\lesssim 44-52$~min, the discs display
$\sim1-5$\,mag outbursts similar to the dwarf nova outbursts observed
in the hydrogen cataclysmic variables
\citep{Levitan2011,Levitan2015,Ramsay2012}. The long period boundary
below which outbursts are observed is not sharp. For example,
Gaia14aae and ASASSN-15fp, with periods of 49.7 and 51.0 min
respectively, were discovered as a result of their outbursts, but the
long-known system GP~Com with $\Porb=46.6$\,min, has never been
observed in outburst. \citet{CannizzoNelemans2015} show that this is a
result of the dependence of the mass transfer rate on the accretor
mass, in the sense that systems with a more massive accretor have a
higher mass transfer rate at a given orbital period. This dependence
is stronger at the long period edge than at the short period edge,
resulting in a mix of outbursting and stable systems near
$\Porb\sim44-52$~min.

\begin{table*}
\caption{The currently known AM~CVn stars ordered by increasing
  period, either the orbital period (most reliably determined from
  spectroscopic observations) or the superhump period (sh) which is
  typically within a few percent longer than the orbital period. (p)
  implies the predicted orbital period based on the outburst
  properties \citep{Levitan2015}. \tickYes\, or \tickNo\, indicates
  whether the source has been seen in outburst. The references and
  full coordinates are given in Table \ref{table1ref}.
\label{sources}}
\centering
\begin{tabular}{lrcrcrrrr}
\hline
Source                   & Period & Outbursts?  & Mag.\,Range  & Reference & Parallax & $\sigma$ & Distance & $\sigma$\\
                         & (mins) &             & (filter)     &           & (mas)    &  (mas)   & (pc)     & (pc)\\
\hline
HM\,Cnc                  & 5.4      &\tickNo  & 21.1       & 1  & & & & \\
V407\,Vul                & 9.5      &\tickNo  & 19.9 (V)     & 2  & 0.095 & 0.327 & 1786 & 667\\
ES\,Cet                  & 10.4     &\tickNo  & 16.5--16.8 & 3  & 0.596 & 0.108 & 1584 & 291 \\
SDSS\,J1351-0643         & 15.7     &\tickNo  & 18.6       & 45 & 0.596 & 0.313 & 1317 & 531\\
AM~CVn                  & 17.1     &\tickNo  & 14.2 & 4 & 3.351 & 0.045 & 299.1 & 4.4\\
SDSS\,J1908+3940         & 18.1     &\tickNo  & 16.1 & 5, 6 & 0.954 & 0.046 & 1044 & 51\\
HP\,Lib                  & 18.4     &\tickNo  & 13.6--13.7 & 7 & 3.62& 0.05 & 276 & 4\\
PTF1\,J1919+4815         & 22.5     &\tickYes & 18.2--21.8 & 8 & 0.550 & 0.327 & 1339 & 555 \\
CX361                    & 22.9     &\tickNo  & 17.6  & 9  & 1.016  & 0.146  & 971  & 156 \\
ASASSN-14cc              & 22.5 (sh)&\tickYes & 16.0--20.0 (V) & 10 & 0.975 & 0.098 & 1019 & 108\\
CR\,Boo                  & 24.5     &\tickYes & 13.8--17.0 & 11 & & & &\\
KL\,Dra                  & 25.0     &\tickYes & 16.0--19.6 & 12 & 1.035 & 0.149 & 956 & 153\\
PTF1\,J2219+3135         & 26.1 (p) &\tickYes & 16.2--20.6 & 14  & & & &\\
V803\,Cen                & 26.6     &\tickYes & 12.8--17.0 & 13, 14 & & & &\\
PTF1\,J0719+4858         & 26.8     &\tickYes & 15.8--19.4 & 15  & 1.144  & 0.301 & 861 & 304 \\ 
ASASSN-15kf              & 27.7 (sh)&\tickYes & 15.0 (V) --19.4 (B) & 16, 17 & & & &\\
YZ~LMi (SDSS\,J0926+3624) & 28.3   &\tickYes & 16.6--19.6 & 18 & 1.824 & 0.549 & 577& 324\\
CP\,Eri                  & 28.4     &\tickYes & 16.2--20.2 & 19 & 0.684 & 0.941 & 964 & 615\\
SDSS\,J1043+5632         & 28.5:(p) &\tickYes & 17.0--20.3 & 14 & 0.830 & 0.668 & 979 & 575\\
CRTS\,J0910-2008         & 29.7 (sh)&\tickYes & 14.0--20.4 (g)      & 47 & 0.695 & 0.477 & 1113 & 561\\
PTF1\,J0943+1029         & 30.4     &\tickYes & 16.9 (R)--20.7 (g) & 20 & & & &\\
CRTS\,J0105+1903         & 31.6     &\tickYes & 16.3-19.6 & 21, 22 & 1.382  & 0.457 & 734 & 374\\
PTF1\,J1632+3511         & 32.7 (p) &\tickYes & 17.9--23.0 & 20 & & & &\\
CRTS\,J0744+3254         & 33: (p)  &\tickYes & 17.4--21.1 & 14 & & & &\\
V406\,Hya                & 33.8     &\tickYes & 14.5--19.7 &  23 & 2.391 & 1.050& 504 & 493\\
PTF1J0435+0029           & 34.3     &\tickYes & 18.4 (R) -- 22.3 (g) & 20 & & & &\\
SDSS\,J1730+5545         & 35.2     &$\sim$ & 18.5 (V) --20.1 & 14, 24a & 1.061 & 0.382& 911 & 420\\
V558\,Vir (2QZ\,J1427-0123) & 36.6 (sh)&\tickYes & 15.0--20.5 & 25 & 1.911 & 1.425& 677 & 595\\
SDSS\,J1240-0159         & 37.4     &\tickYes & 16.8--19.8 & 27  & 1.857 & 0.611& 577 & 365\\
NSV1440                  & 37.5 (sh)&\tickYes & 12.4(V) -- 17.9(G) & 47 & 2.971 & 0.142 &  337 & 17 \\
V744\,And (SDSS\,J0129+3842) & 37.6 &\tickYes & 14.5--19.8 & 14, 26, 29 & 2.066 & 0.529 & 508& 239\\
SDSS\,J1721+2733         & 38.1     &\tickYes & 16.0--20.1 & 14, 31 & 0.798 & 0.665 & 995 & 578\\  
ASASSN-14mv              & 41: (sh) &\tickYes & 11.8 (V) -- 18.1 (V) & 16, 17, 46 & 4.069 & 0.119 & 247 & 7\\
ASASSN-14ei              & 43: (sh) &\tickYes & 11.9--17.6 (B) & 17, 33 & 3.92 & 0.045& 255 & 4\\
SDSS\,J1525+3600         & 44.3     &\tickNo  & 20.2 & 31  & 1.928  & 0.276  & 524  & 90\\
SDSS\,J0804+1616         & 44.5     &\tickYes & 17.8--19.0 & 28, 30 & 1.203 & 0.210 & 828 & 173\\
SDSS\,J1411+4812         & 46.0     &\tickYes  & 19.4--19.7 & 29, 30, 50 & 2.361 & 0.305 & 429 & 65\\
GP\,Com                  & 46.6     &\tickNo  & 15.9--16.3 & 32 & 13.731 & 0.060 & 73.0 & 0.4\\
CRTS\,J0450-0931         & 47.3 (sh)&\tickYes & 14.8--20.5 & 34 & & & &\\
SDSS\,J0902+3819         & 48.3     &\tickYes & 13.8 (V) -- 20.2 (g) & 15, 31, 35 & 2.519 & 0.936  & 461 & 435\\
Gaia14aae                & 49.7     &\tickYes & 13.6 (V) -- 18.7 (g) & 36, 51 & 3.871  & 0.155 & 259 & 11\\
ASASSN-17fp              & 51.0 (sh)&\tickYes & 15.7--20+ & 37  & & & &\\
SDSS\,J1208+3550         & 53.0     &\tickNo  & 18.9--19.4 & 30, 39, 40 & 5.005 & 0.416 & 202& 18\\
SDSS\,J1642+1934         & 54.2     &\tickNo  & 20.3 & 31, 40 & 0.621 & 0.730 & 1044  & 604\\
SDSS\,J1552+3201         & 56.3     &\tickNo  & 20.2--20.6 & 30, 42 & 2.395 & 0.609 & 443& 227\\
SDSS\,J1137+4054         & 59.6:    &\tickNo  & 19.0 & 24b & 4.838& 0.310 &209 &14\\
V396\,Hya                & 65.1     &\tickNo  & 17.6  & 44 & 10.694 & 0.148 & 93.6 & 1.4\\
SDSS\,J1319+5915         & 65.6     &\tickNo  & 19.1 & 41,49 & 4.894 & 0.240 & 205 & 10\\
PU\,Aqr (SDSS\,J2047+0008) & ?      &\tickYes & 17.0--24 & 39 & & & &\\ 
CRTS\,J0844-0128         & ?        &\tickYes & 17.4-20.3 & 14 & 0.395 & 0.341 & 1474 & 597\\
PTF1\,J0857+0729         & ?        &\tickYes & 18.6-21.8 & 14  & & & &\\
SDSS\,J1505+0659         & ?        &\tickNo  & 19.1 & 24b & 6.299 & 0.453 & 160 & 12\\
PTF1\,J1523+1845         & ?        &\tickYes &   17.6--23.5 & 14 & & & &\\
ASASSN-14fv              & ?        &\tickYes & 14.6 (V) -- 20.5 (B) & 42 & & & &\\
Gaia16all                & ?        &\tickYes & 16.2 -- 20.6(G)    & 48 & 0.760 & 0.859 & 956 & 605 \\
SDSS\,J0807+4852         & ?        &\tickNo  & 19.5(V)--20.4(g) & 52 & 0.055 & 2.42 & 883 & 648 \\
\hline
\end{tabular}
\end{table*}

\begin{figure*}
\includegraphics[width=0.48\textwidth]{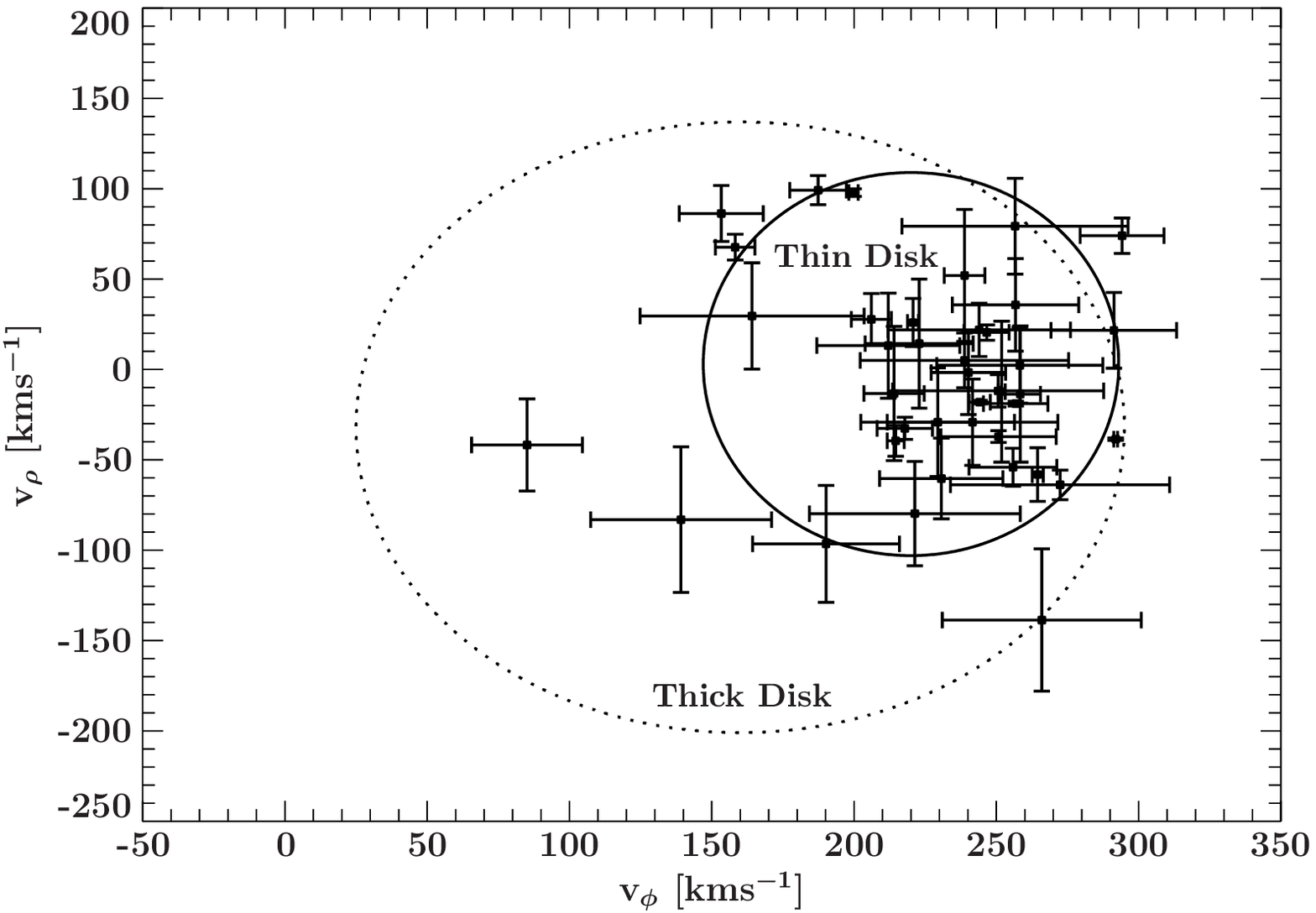}\hfill
\includegraphics[width=0.48\textwidth]{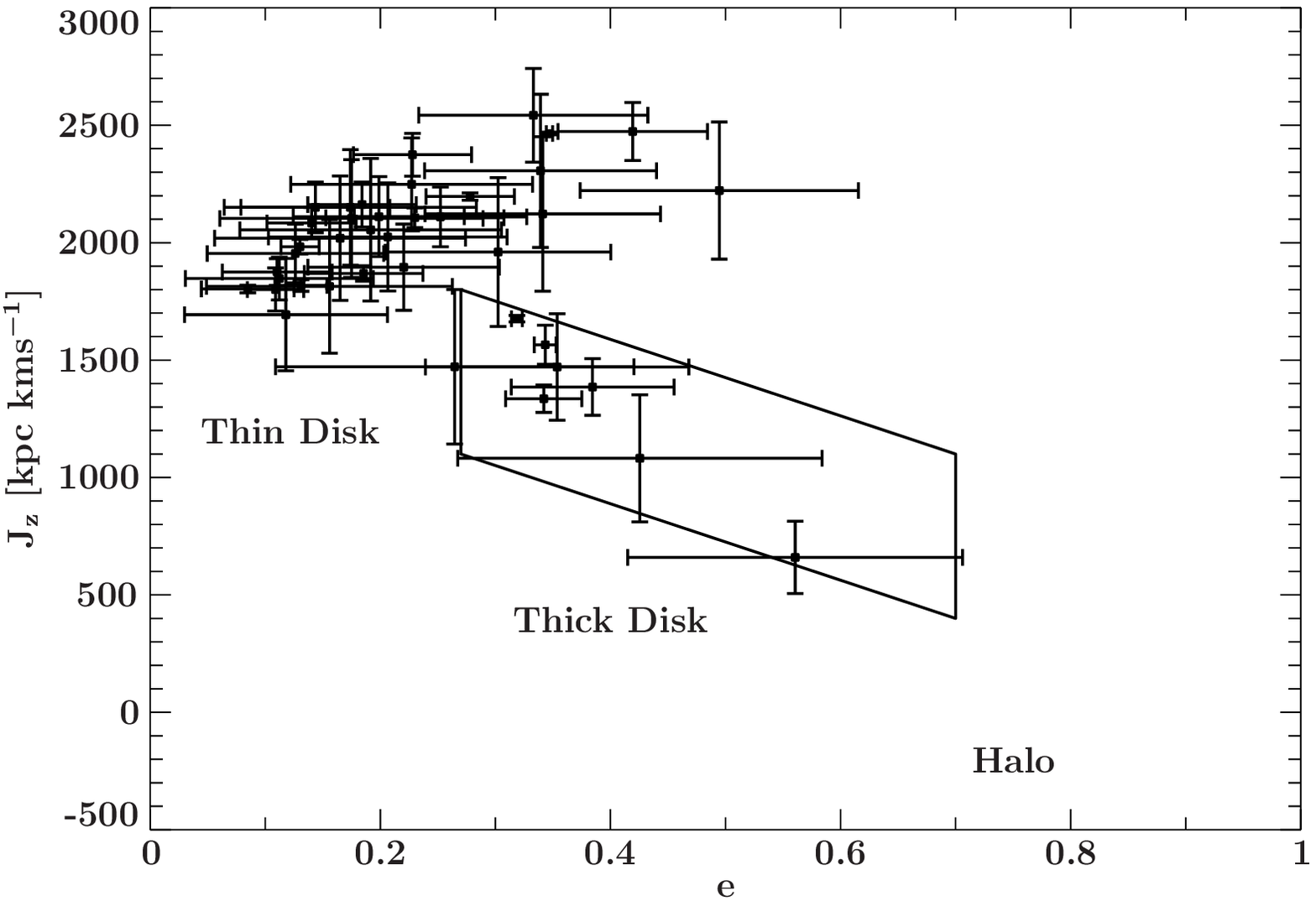}\\
\vspace{0.2cm}
\caption{The AM~CVn stars in the plane of the velocity in the
  direction of Galactic rotation ($V_\phi$) and the Galactic center
  ($V_\rho$) (left hand panel) and the orbital eccentricity ($e$) and
  the angular momentum in the Galactic $z$ direction($J_{z}$) (right
  hand panel). The solid and dotted ellipses render the 3$\sigma$ thin
  and thick disk contours in the $V_\phi$--$V_\rho$ diagram, while the
  solid box in the $e$--$J_{z}$ marks the thick disk region as
  specified by \citet{Pauli06}.}
\label{fig:galkin}
\end{figure*}

\begin{table}
\caption{\Gaia\, DR2 and \HST\, parallaxes of the five AM~CVn stars
  which were measured using the \HST\, Fine Guidance
  Sensor. \label{hstversusgaia}} \centering
  \begin{tabular}{lcc}
    \hline
    ~          & \HST      & \Gaia     \\
    Source     & parallax  & parallax \\
    ~          & (mas)     & (mas)    \\
\hline
AM~CVn    & $1.65\pm0.30$  & $3.351\pm0.045$ \\
HP\,Lib    & $5.07\pm0.33$  & $3.622\pm0.053$  \\
CR\,Boo    & $2.97\pm0.34$  & N/A  \\
V803\,Cen  & $2.88\pm0.24$  & N/A  \\
GP\,Com    & $13.34\pm0.33$ & $13.73\pm0.06$ \\
\hline
\end{tabular}
\end{table}

\section{Parallaxes and distances of AM~CVns in Gaia DR2} \label{sec:distances}

In Table~\ref{sources} we show the parallaxes for all AM~CVn stars
included in the \Gaia\, DR2: out of the 56 known systems 41 have
parallax measurements. The closest system is the long known GP~Com
with a parallax of 13.73$\pm$0.06 mas, with the most distant being
V407~Vul which has a parallax consistent with zero (0.10$\pm$0.33
mas). In Table \ref{hstversusgaia} we compare the parallaxes for three
sources which have both \Gaia\, and \HST\, parallaxes (V803~Cen and
CR~Boo have no parallax in \Gaia\, DR2). There is reasonable agreement
for GP~Com. However, HP~Lib is significantly more distant and AM~CVn
significantly closer compared to the results of \HST. Indeed the
\HST\, parallax derived for AM~CVn implied a greater distance than
other estimates and therefore implied a high luminosity and mass
accretion rate, all of which were problematic in explaining.

We convert parallaxes from the \Gaia\, DR2 into distances following
the guidelines from \citet{Bailer-Jones2015,Astraatmagja2016} and
\citet{Luri2018}.  Given the measured parallax $\varpi$ with the
uncertainty $\sigma_{\varpi}$, the probability density of possible
values for the distance can be found by using Bayes' theorem:
\begin{equation}
P(r|\varpi,\sigma_{\varpi})= \frac{1}{Z}\  P(\varpi|r,\sigma_{\varpi})\  P(r),
\end{equation}
where $r$ is the distance, $P(\varpi|r,\sigma_{\varpi})$ is the
likelihood function, that can be assumed Gaussian
\citep{Lindegren2018}. $P(r)$ is the prior and $Z$ is a normalisation
constant. The prior is an arbitrary function that expresses our
knowledge on the distribution of the distances of AM~CVn stars and
allows us to introduce assumptions in the inference procedure that are
not related to the measurement of the parallax itself.  The properties
of various priors and their performance on the data from the \Gaia\,
DR1 have been investigated in \citet{Astraatmagja2016}.  For this work
we adopt the exponentially decreasing space density prior:
\begin{equation}  \label{eqn:prior}
P(r) = \begin{cases}\frac{ r^2}{2L^3} \exp(-r/L) & \text {if} \  r > 0,\\
0 & \text{otherwise}, \end{cases}
\end{equation}
where $L$ is the scale length.  This prior implies a constant space
density of AM~CVn stars for $r << L$ and an exponential drop for $r >>
2L$, where $2L$ corresponds to the peak of the distribution. The
choice of the value for $L$ needs to be fine-tuned to reproduce the
distribution of AM~CVn stars with the distance. We adopt $L=400\,$pc
calibrated on the mock population of double white dwarf binaries
(progenitor systems of AM~CVns) from \citet{Korol2017}. For more
detailed argumentation we refer the reader to \cite{Kupfer2018}.  In
Table \ref{sources} we show the inferred distance and uncertainty
(which covers the 90 percent confidence interval) for those AM~CVn
systems which have parallax measurements in the \Gaia\, DR2. The
distances range from 73~pc (GP~Com) to 1.8~kpc (V407~Vul), although
the latter is rather uncertain. The median distance is 580~pc. For
sources closer than 340~pc the mean uncertainty on the distance is
9~pc, and for those between 500--1000~pc it is 380~pc.

\section{Galactic distribution}
\label{sec:galdist}

For the 41 systems with parallax and proper motions from \Gaia\, DR2
(shown in Table \ref{population}), we calculate 3D kinematics and put
constraints on their population membership: thin disk, thick disk, and
halo. Only AM~CVn, SDSSJ\,1908+3940, CP\,Eri, SDSS\,J1730+5545 and
SDSS\,J1240-0159 have measured systemic velocities based on radial
velocity curves of their accretion disc lines. Seven additional
systems have a strong central spike feature which can be used to
measure the radial velocity. These central spike lines are believed to
originate close to the photosphere of the accretor and are shifted
with the gravitational redshift of the accretor. Assuming an accretor
with $M=0.8$\,\Msun\, and a radius of $R=0.01$\,\Rsun\, leads to a
gravitational redshift of $50$~km\,s$^{-1}$, so we correct the
measured velocity from the central spike by $50$~km\,s$^{-1}$. For the
remaining systems with no measured radial velocity, we assume
0~km\,s$^{-1}$ with an error of $\pm50$~km\,s$^{-1}$.

Combined with right ascension and declination, we calculate velocity
in the direction of the Galactic Centre ($V_\rho$) and the Galactic
rotation direction ($V_\phi$), the Galactic orbital eccentricity
($e$), and the angular momentum in the Galactic $z$ direction
($J_z$). The Galactic radial velocity $V_\rho$ is negative towards the
Galactic centre, while stars that are revolving on retrograde orbits
around the Galactic Centre have negative $V_\phi$. Stars on retrograde
orbits have positive $J_z$. Thin disk stars generally have very low
eccentricities $e$. Population membership can be derived from the
position in the $V_\rho$ - $V_\phi$ diagram and the $J_z$ - $e$
diagram following the description in \citet{Pauli03,Pauli06}.

We show the results in Figure \ref{fig:galkin}. We find that the most
of the systems show a Galactic orbit typical for a thin disc
population (see Table \ref{population}). About 10 systems have a
typical thick disc orbit. None of the systems show a halo orbit.  In a
small number of AM~CVn stars, the extreme depletion of heavy elements
have been taken as evidence that these stars were halo objects
(GP~Com: \citet{Marsh1991}; V396~Hya: \citet{Nagel2009} and
PTF1\,J0719+4858: \citet{Gehron2014}. Our study has shown that these
AM~CVn stars are likely thin or thick disk objects and not halo
objects. It is of interest to understand how disk objects can have
such low abundances of heavy elements.

\section{Determining the line of sight extinction}
\label{sec:ext}

To determine the absolute magnitude of AM~CVn stars we need to
subtract the effects of interstellar extinction. Although 47 of the 56
AM~CVn stars shown in Table \ref{sources} lie at Galactic latitudes
$|b|>20^{\circ}$ (implying the extinction is likely to be low), we
have determined the line-of-sight extinction to our sources using
3D-dust maps derived from Pan-STARRS1 data \citep{Green2018b}. For
each AM~CVn star, we derived the extinction, $E_{B-V}$, for the sky
co-ordinates and distance of that star given in Table \ref{sources}
(the uncertainty is typically $E_{B-V}\sim$0.02). To determine the
reddening, we use $A_{V}=R \times E_{B-V}$, where we assume the
standard value of $R=$3.2. To obtain the reddening in the $g$ band, we
assume $A{g}=1.1 \times A_{V}$ \citep{Cardelli1989}. For those AM~CVn
stars at declinations too south to feature in the Pan-STARRS1
catalogue we use the dustmaps of \citep{Schlafly2011} which give the
extinction to the edge of the Galaxy in that line of sight. (This
  step was required for only two stars and given that the upper limits
  are only 0.2 mag, we do not expect that this uncertainty will
  significantly effect their place on the period -- absolute magnitude
  relationship).

We show the reddening to those AM~CVn stars with parallaxes in Table
\ref{ext}. The median reddening of our sample is $A_{g}$=0.14 mag,
with 90 percent of sources having $A_{g}<$0.35 mag. Therefore for the
majority of our targets the effects of interstellar absorption has a
small effect on the resulting absolute magnitudes. The only source
with a high extinction, V407~Vul, was previously known to have a high
degree of reddening \citep{Motch1996}.

\section{The absolute magnitudes of AM~CVn stars}
\label{abmag}

\begin{figure}
\begin{center}
\setlength{\unitlength}{1cm}
\begin{picture}(16,6.5)
\put(0,-0.2){\includegraphics{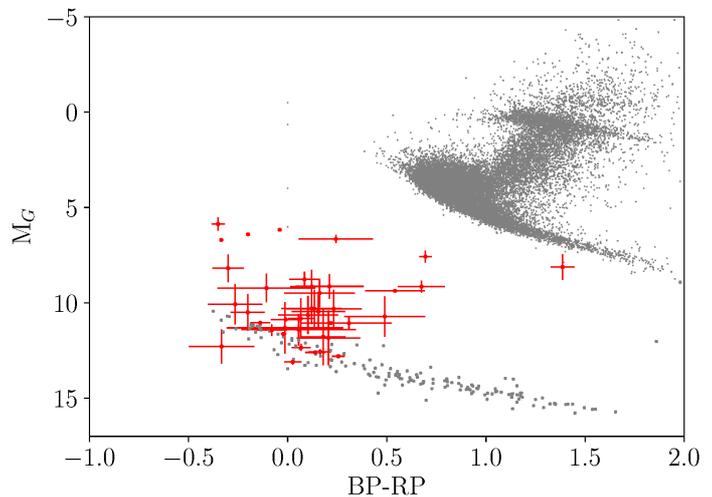}}
\end{picture}
\end{center}
\caption{An HR diagram made using {\sl Gaia} DR2 data. In red we show
  the absolute $G$ magnitude, $M{_G}$, as a function of the blue-red
  colour, BP-RP, for AM~CVns with known parallaxes.  No reddening or
  extinction corrections have been applied to this figure.  For
  comparison, in grey points we show a sample of essentially randomly
  selected main sequence stars which have a parallax better than
  1~percent. For single white dwarfs shown as slightly larger grey
  points) we have used the sample of white dwarfs within 20~pc
  \citep{Hollands2018}. The majority of AM~CVns appear to be brighter
  than the single white dwarf track, though some long-period systems
  lie on the track. A clear outlier is V407~Vul, which has BP-RP
  ~$\approx 1.5$ through a combination of severe reddening and
  contamination from an unresolved nearby G-star.  }
\label{hrdiag} 
\end{figure}

Using the distances shown in Table \ref{sources}, we show in Figure
\ref{hrdiag} the absolute magnitude, $M_{G}$, against the BP-RP (the
`Blue Photometer' and `Red Photometer' colour derived from {\sl Gaia}
data) of AM~CVn stars and a sample of Galactic stars with accurate
parallax measurements. The single white dwarf track can be seen in the
lower left of the figure. High state AM~CVn stars are significantly
brighter than the isolated white dwarfs, being $M_{G}\sim$6. The
majority of quiescent AM~CVn stars appear 1--2 mag brighter than the
white dwarfs, being $M_{G}\sim$8--12, though a minority lie on the
white dwarf track.

As the wavelength range of the {\sl Gaia} $G$ band filter is very
broad, it is not the optimal filter with which to test evolutionary
models which usually predict the absolute magnitude in the $V$ or $g$
bands. We have therefore collated all the available multi-filter
photometry of the AM~CVn stars from the Pan-STARSS1, Skymapper,
\galex\, and SDSS surveys and we show these in
Appendix~\ref{Galex}--\ref{Skymapper} and associated tables. Since
most surveys release mean or median values for each source and filter,
the effects of outbursts are generally smoothed out and usually they
reflect the quiescent brightness for outbursting systems. In Table
\ref{ext} we show the mean quiescent $g$ mag which we derive for these
systems. For most systems this magnitude is found from analysis of the
$g$-band data shown in the Appendix. In the case of ASASSN-14cc, a
short-period system which is in outburst for around 60 percent of the
time, the magnitude measured by Skymapper is representative of its
outburst magnitude. We therefore take its quiescent magnitude from
\citet{Kato2015}.

\begin{figure}
\begin{center}
\setlength{\unitlength}{1cm}
\begin{picture}(16,6.5)
\put(0.2,-0.4){\includegraphics{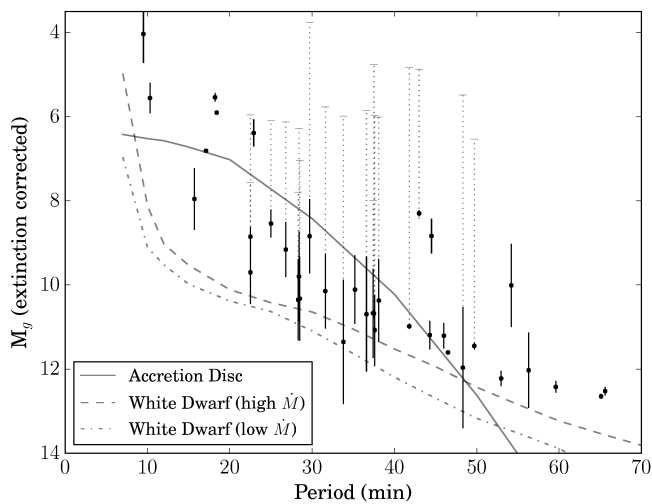}}
\end{picture}
\end{center}
\caption{The extinction-corrected absolute magnitudes ($M_{g}$) of
  AM~CVn binaries as a function of orbital period.  For known
  outbursting systems, we plot the quiescent magnitude, and show the
  magnitude change during outburst as a dotted line.  The dashed line
  is the predicted $M_{V}$ of the accretion disk as originally taken
  from \citet{Nelemans2004} while the solid line shows the evolution
  of a 0.65\Msun accreting white dwarf with a high-entropy white dwarf
  donor, including accretion heating, from \citet{Bildsten2006}.  Most
  systems appear somewhat brighter than an isolated white dwarf model
  during quiescence, and close to the duty-cycle-scaled disc magnitude
  during outburst.  }
\label{PorbMv} 
\end{figure}

Based on these quiescent magnitudes, we calculate the absolute
magnitudes shown in Table~\ref{ext}.  In Figure~\ref{PorbMv} we plot
the absolute magnitudes of AM~CVn binaries as a function of orbital
period, for the 37 systems in which both magnitude and orbital period
are known. For those systems which outburst we also plot an
approximate outburst magnitude.  For comparison, we show the predicted
magnitudes of the central white dwarf and the accretion disc, both of
which vary in magnitude as a function of orbital period.  We show two
cooling tracks for a 0.65\,\Msun\, central white dwarf, both taken
from \citet{Bildsten2006}: one assumes a low mass transfer rate (as
would be seen for a cool, low-mass white dwarf donor) and one assumes
a high mass transfer rate (as would be seen in a hot, high-mass dwarf
donor). The donor masses assumed span the range of white dwarf donor
masses predicted by \citet{Deloye2005}, but note that higher-mass
donors are possible through other evolutionary channels.  The
accretion disc magnitudes are calculated by \citet{Nelemans2004}.

As Figure~\ref{PorbMv} shows, a strong decrease in AM~CVn absolute
magnitude with increasing orbital period is seen. High-state systems
($\Porb \lesssim 20$min) lie close to the predicted accretion disc
magnitudes, as expected for these disc-dominated systems. Quiescent
systems lie a little above the white dwarf cooling track, with an
excess that generally decreases with increasing period, though there
is considerable scatter between systems. For outbursting systems, the
outburst magnitude in most cases is far in excess of the mean disc
magnitude.

All AM~CVn systems shown in Figure~\ref{PorbMv} are brighter than the
predicted white dwarf cooling track for a high-mass white dwarf donor,
and significantly brighter than the cooling track for a low-mass white
dwarf donor. For systems at $\Porb > 50$\,min, which are dominated by
their white dwarf magnitudes, this is particularly interesting. It may
suggest that AM~CVn donors have a systematically higher mass than
predicted. Measurements of AM~CVn mass ratios imply a similar trend
\citep{Green2018a}. If true, this would imply either that white dwarf
donors have higher temperatures (and hence higher masses for a given
orbital period) than predicted, or that the majority of systems have
non-white dwarf donors.

Several individual systems are worthy of note:
\begin{itemize}
\item V407~Vul appears unusually bright due to the G star which is co-aligned on the sky. \citet{Barros2007}
  estimated that the G star contributes 92\% of the light in the \textit{g}-band, giving a magnitude difference of 2.7.
\item SDSS~J1351-0643 has an unexpectedly low magnitude for a
  high-state system, though its uncertainties are large. This low
  magnitude could be explained if the system is at a high inclination,
  such that the disc is viewed close to edge-on. Its broad
  double-peaked spectrum and photometrically-visible disc precession
  also support a high inclination.
\item ASASSN-14ei, SDSS\,J0804+1616, Gaia14aae, and SDSS\,J1642+1934
  seem to be unusually bright for their orbital period. For some of
  these systems there is evidence they may have a high mass transfer
  rate: SDSS\,J0804+1616 is unusually variable \citep{Ramsay2012},
  while Gaia14aae is among the longest-period outbursting AM~CVns and
  has an unexpectedly high-mass donor star
  \citep{Campbell2015,Green2018_14aae}. As discussed in
  Section~\ref{sec:components}, the elevated magnitude of Gaia14aae is
  almost entirely due to its hot central white dwarf.
\item Several systems have outburst magnitudes that are lower than the
  general trend, including PTF1\,J1919+4815, YZ~LMi, and
  SDSS\,J1240-0159. The first two of these systems are partial
  eclipsers and therefore high-inclination, which will reduce the
  observed outburst magnitude. The reason for the weak outburst of
  SDSS\,J1240-0159, reported by \citet{RoelofsPhD}, is uncertain.
\end{itemize}

\begin{table}
\caption{We indicate the reddening in the $g$ band ($A_{g}$), which
  have been determined using Pan-STARRS1 data
  \citep{Green2018b}. Those stars where the reddening has been
  estimated using the dust maps of \citep{Schlafly2011} have a
  superscript $m$ to indicate it is the reddening to the edge of the
  Galaxy. The third column shows the quiescent $g$ magnitude as
  determined from the survey magnitudes outlined in the Appendix. In
  the final column we show the absolute magnitude, $M_{g}$.  These
  data are also included in the online table. \label{ext}}
\begin{center}
  \begin{tabular}{lrrrr}
    \hline
    Source   & $P_{orb}$ & $A_{g}$ & $g_{q}$ & $M_{g}$ \\
              & (min)    & (mag)   & (mag) & (mag) \\
\hline
V407 Vul & 9.5 & 4.9 & 20.16 & 4.0$\pm$0.7\\
ES Cet & 10.4 & 0.1 & 16.66 & 5.6$\pm$0.4\\
SDSS\,J1351-0643 & 15.7 & 0.1 & 18.66 & 8.0$\pm$0.7\\
AM~CVn & 17.1 & 0.07 & 14.26 & 6.81$\pm$0.03\\
SDSS\,J1908+3940 & 18.1 & 0.53 & 16.16 & 5.5$\pm$0.1\\
HP Lib & 18.4 & 0.49 & 13.60 & 5.91$\pm$0.03\\
PTF1\,J1919+4815 & 22.5 & 0.18 & 20.51 & 9.7$\pm$0.8\\
CX361 & 22.9 & 1.16 & 17.48 & 6.4$\pm$0.3\\
ASASSN-14cc & 22.5 & 0.12 & 19.0 & 8.9$\pm$0.2\\
KL~Dra & 25.0 & 0.18 & 18.62 & 8.5$\pm$0.3\\
PTF1\,J0719+4858 & 26.8 & 0.32 & 19.15 & 9.2$\pm$0.7\\
YZ LMi & 28.3 & 0.07 & 19.23 & 10.4$\pm$1.0\\
CP Eri & 28.4 & 0.35 & 20.07 & 9.8$\pm$1.1\\
SDSS J1043+5632 & 28.5 & 0.07 & 20.35 & 10.33$\pm$1.0\\
CRTS J0910-2008 & 30.2 & 0.32 & 19.39 & 9.9$\pm$0.9\\
CRTS J0105+1903 & 31.6 & 0.18 & 19.65 & 10.1$\pm$0.9\\
V406 Hya & 33.8 & 0.11 & 19.97 & 11.4$\pm$1.5\\
SDSS\,J1730+5545 & 35.2 & 0.14 & 20.05 & 10.1$\pm$0.8\\
NSV1440 & 36.3 & 0.21$^{m}$ & 18.53  & 10.7$\pm$0.1\\
V558 Vir & 36.6 & 0.14 & 19.99 & 10.7$\pm$1.4\\
SDSS\,J1240-0159 & 37.4 & 0.07 & 19.55 & 10.7$\pm$1.1\\
V744 And & 37.6 & 0.28 & 19.88 & 11.1$\pm$0.8\\
SDSS\,J1721+2733 & 38.1 & 0.07 & 20.43 & 10.4$\pm$1.0\\
ASASSN-14mv & 41 & 0.04 & 17.98 & 11.0$\pm$0.1\\
ASASSN-14ei & 43 & 0.21$^{m}$ & 15.54 & 8.94$\pm$0.03\\
SDSS\,J1525+3600 & 44.3 & 0.07 & 19.86 & 11.2$\pm$0.3\\
SDSS\,J0804+1616 & 44.5 & 0.18 & 18.60 & 8.8$\pm$0.4\\
SDSS\,J1411+4812 & 46.0 & 0.07 & 19.44 & 11.2$\pm$0.3\\
GP~Com & 46.6 & 0.00 & 15.92 & 11.60$\pm$0.01\\
SDSS\,J0902+3819 & 48.3 & 0.11 & 20.39 & 12.0$\pm$1.4\\
Gaia14aae & 49.7 & 0.04 & 18.55 & 11.4$\pm$0.1\\
SDSS\,J1208+3550 & 53.0 & 0.07 & 18.82 & 12.2$\pm$0.2\\
SDSS\,J1642+1934 & 54.2 & 0.25 & 20.35 & 10.0$\pm$1.0\\
SDSS\,J1552+3201 & 56.3 & 0.14 & 20.40 & 12.0$\pm$0.9\\
SDSS\,J1137+4054 & 59.6 & 0.07 & 19.09 & 12.4$\pm$0.1\\
V396 Hya & 65.1 & 0.18 & 17.68 & 12.64$\pm$0.03\\
SDSS\,J1319+5915 & 65.6 & 0.04 & 19.12 & 12.6$\pm$0.1\\
\hline
\end{tabular}
  \end{center}
\end{table}

\section{Component magnitudes of eclipsing AM~CVns}
\label{sec:components}

Three AM~CVns are currently known which undergo eclipses. In Gaia14aae
and YZ~LMi, the central white dwarf is eclipsed, while in
PTFJ~1919+4815, only the edge of the disc is eclipsed. The first two
systems present an opportunity to measure the flux contributions from,
and hence absolute magnitudes of, their central white dwarfs. We can
additionally measure the flux contribution from the `bright spot'
component of these systems, this being a bright region located at the
point of collision between the accretion disc and the stream of
infalling matter from the donor.

Photometry of YZ~LMi in Sloan \textit{ugr} bands and of Gaia14aae in
\textit{ugri} bands were presented in \citet{Copperwheat2011} and
\citet{Green2018_14aae} respectively. Both papers quote the measured
white dwarf fluxes, from which we calculate the absolute
magnitudes. To calculate the bright spot contributions, we obtained
the best-fit models presented in both papers and measured the bright
spot contribution directly. Uncertainties were calculated from the
1-$\sigma$ spread of the MCMC results created for those papers when
converging on the data. The absolute magnitudes we calculate for these
components are presented in Table~\ref{tab:components}. In the case of
YZ~LMi, large uncertainties result from the poorly constrained {\sl
  Gaia} parallax of the system. The \textit{i}-band Gaia14aae
magnitudes are poorly constrained due to the small number of eclipses
observed in that band.

\subsection{Central white dwarfs}

These central white dwarf magnitudes may be compared with the
predicted white dwarf magnitude track shown in Figure~\ref{PorbMv},
for corresponding orbital periods.  It should be noted that both white
dwarfs are somewhat higher mass ($\approx 0.85$\,\Msun) than the mass
assumed in that model ($0.65$\,\Msun). A larger mass effects the
magnitude in two ways: the reduction in size of the white dwarf causes
it to appear fainter, but the smaller surface area increases the
effect of accretion heating on temperature.  With this caveat, the
central white dwarf \textit{g} magnitude of YZ~LMi agrees reasonably
well with the predicted value of 10.6.  For Gaia14aae, the central
white dwarf magnitude of 11.66(9) is considerably brighter than the
predicted magnitude of 12.4.  As shown in \citet{Bildsten2006}, the
white dwarf mass is less significant than accretion rate for
long-period systems like Gaia14aae.  It therefore seems likely that
the elevation of the central white dwarf magnitude of Gaia14aae above
the model track is due to a higher accretion rate than predicted,
resulting in a hotter white dwarf.

Given the tightly constrained absolute magnitudes of Gaia14aae, the
temperature of the central white dwarf can be estimated. From DB
atmosphere models \citep{Bergeron2011, Tremblay2011, Kowalski2006,
  Holberg2006}, for a surface gravity $\log (g) \approx 8.5$ as
measured by \citet{Green2018_14aae}, these magnitudes in all four
colour bands predict a temperature of $\approx 17000 \pm 1000$~K. This
temperature disagrees with the previously established temperature of
$12900 \pm 200$~K established from UV flux \citep{Campbell2015}. The
discrepancy may be the result of metals accreted from the donor star,
in particular nitrogen, which are expected given the evolved nature of
the donor \citep{Nelemans2010}. Such metals would cause absorption in
the UV not present in a pure DB atmosphere, making the UV-derived
temperature unreliable. We therefore interpret 17000~K as the most
likely temperature of the white dwarf.

\subsection{Bright spots}
\label{brightspot}

The magnitudes of the bright spots in these systems allow for their
instantaneous mass transfer rates, $\dot{M}$, to be estimated. If the
luminosity of the bright spot is equal to the kinetic energy released
by the infalling matter as it slows to match the disc velocity, the
luminosity can be described as
\begin{equation}
L = \frac{1}{2} \dot{M} \lvert \vec{V}_\mathrm{stream} - \vec{V}_\mathrm{disc} \rvert ^2
\label{eq:masstransfer}
\end{equation}
where $\vec{V}_\mathrm{stream}$ and $\vec{V}_\mathrm{disc}$ are
vectors describing the velocities of material in the stream and disc
at the point of intersection between the two. Assuming that the disc
material follows a Keplerian orbit and that the stream trajectory is
ballistic, these can be calculated based on the measured stellar
masses of the system. The luminosity found by
Equation~\ref{eq:masstransfer} can then be converted to a magnitude
using an assumed spectral response for the bright spot. We assume the
spectral response of a 12000~K blackbody, as was observed for the
cataclysmic variable IP~Peg \citep{Marsh1988}.  To account for our
uncertainty in this choice of spectral response, we increase the
uncertainties on the predicted magnitude by 0.2~mag, and propagate
this through to the uncertainties on $\dot{M}$.  We find $\log
(\dot{M} / M_\odot \mathrm{yr}^{-1}) = -10.6 \pm 0.4$ for YZ~LMi and
$-10.74 \pm 0.07$ for Gaia14aae.

For comparison we calculate theoretical values of $\dot{M}$ based on
the photometrically measured masses, using the relations in
\citet{Deloye2007}. It is necessary to assume a value for the donor
star's response to mass loss, $d\log(R)/d\log(M)$. For this we assume
0.2, which approximates the tracks in \citet{Deloye2007} for donors
evolving isothermally.  For YZ~LMi, the predicted $\log (\dot{M} /
M_\odot \mathrm{yr}^{-1}) = 10.0$ is within 1.5-$\sigma$ of our
measured value. Given that these uncertainties are likely to be
non-Gaussian, we do not consider this discrepancy to be significant.
In the case of Gaia14aae, an assumed donor response of 0.2 implies
$\log (\dot{M} / M_\odot \mathrm{yr}^{-1}) = 10.77$, in good agreement
with our measured mass transfer rate.

The agreement in the case of Gaia14aae is, to some extent,
surprising. \citet{Deloye2007} predict $d\log(R)/d\log(M) = 0.2$ for
AM~CVn donors during their short-period evolution. However, once they
evolve to periods $\gtrsim 40$~min, a change of state in the donor is
expected that would result in this value decreasing. The fact that 0.2
still appears to hold for Gaia14aae agrees with our suggestion in
\citet{Green2018_14aae} that the donor has not yet undergone this
change of state.

Recent work by \citet{Piersanti2015} predict the mass accretion
  rate for systems with periods shorter than 22 min. However, we use
  the work of \citet{Bildsten2006} who give a relation between the
  mass transfer rate and the temperature of the accretion-heated
  central white dwarf in AM~CVn systems for the entire orbital period
  seen in AM CVn stars. \citet{Bildsten2006} predict that,
for long-period ($P \gtrsim 30$~min) and hence low-$\dot{M}$ systems,
energy radiated from the white dwarf core will heat the surface more
than energy from accretion. The surface temperature will then be
higher than would be predicted from $\dot{M}$ alone. This can be seen
in Gaia14aae, where the $\dot{M}$ measured above would suggest a
blackbody white dwarf temperature of 5500K. The difference between
this and our measured temperature of 17000K implies, as predicted,
that the temperature has decoupled from $\dot{M}$.

\begin{table}
\caption{Absolute magnitudes of the central white dwarfs and bright spots in two eclipsing systems: YZ~LMi and Gaia14aae. 
\label{tab:components}}
\begin{center}
  \begin{tabular}{lccc}
    \hline
    Source   & Filter & White Dwarf & Bright Spot \\
\hline
YZ~LMi & \textit{r} & 10.9(1.0) & 14.7(1.6)\\
 & \textit{g} & 11.2(1.0) & 14.1(1.5) \\
 & \textit{u} & 10.7(1.0) & 14.5(1.6) \\
 \hline
Gaia14aae & \textit{i} & 12.17(10) & 15.7(1.1) \\
 & \textit{r} & 11.95(9) & 14.6(2)\\
 & \textit{g} & 11.66(9) & 14.8(2)\\
 & \textit{u} & 11.58(10) & 14.4(7) \\
\hline
\end{tabular}
  \end{center}
\end{table}

\section{Mass transfer rate as a function of orbital period}
\label{rate}

One of the reasons that determining the theoretical space density of
AM~CVn stars has so many uncertainties, is that although there are
three likely formation channels, the relative importance of each channel is not well determined. The formation
channels largely differ in the nature of the mass donating star. A
white dwarf donor would start mass transfer at short orbital periods,
whilst the helium star, or highly evolved CV donor, would initially
evolve to short orbital periods before evolving to long periods as
the mass donating star becomes fully degenerate \citep[see][for
a detailed overview]{Solheim2010}.

If we were able to accurately determine the mass transfer rate for
many sources with a range of orbital period we would make progress in
understanding the relative importance of the formation channels since
they follow different tracks on the $\Porb-\dot{M}$ plane. Determining
reliable values for $\dot{M}$ requires photometric information across
a wide range of the spectrum, including the UV, where much of the
accretion luminosity is emitted (especially for short period systems,
e.g. \citet{Ramsay2005}, \citet{Ramsay2006}.

For AM~CVn stars which have a parallax accurate to within 20 percent,
we converted the multi-band photometric data outlined in the appendix
to flux units (Jy). (We also included SDSS J1351 since it has a short
orbital period -- 15.7 min -- and is therefore interesting from an
evolutionary point of view). Some systems, such as KL Dra and AM CVn
show a large scatter between the multi-filter flux measurements
(possibly due to outbursts in the case of KL Dra, and perhaps
  significant flickering in the case of AM CVn) and were therefore
not suited to determining their mass transfer rate by this
means. There were 15 systems which had suitable data and we show their
spectral energy distributions in Figure ~\ref{sed1} and Figure
~\ref{sed2}. For the shortest orbital period systems, the energy
distribution is generally still increasing at our bluest flux point
(\galex\, FUV). For long period systems we can locate the wavelength
which the flux appears to peak.

To determine the mass transfer rate we sum up the contribution of flux
emitted from the white dwarf and accretion disk. Although the
  cooling models of \citet{Bildsten2006} assume that the primary star
  has a mass of 0.65 \Msun, the primary in both of the eclipsing
  systems is $M_{1}$=0.8 \Msun. We therefore set $M_{1}=0.8\pm$0.1
  \Msun\ for the non-eclipsing systems. We calculate the radius of
the primary using the formula of Eggleton's quoted in
\citet{VerbuntRappaport88}. We also assume that the disc is accreting
at a steady state and has a number of annuli (40) which we compute the
temperature assuming a black body. Again apart from Gaia14aae the
system inclination is not well determined and therefore we fix
cos(i)=0.5 (apart from Gaia14aae where $i=86.3^{\circ}$). An
additional parameter was the radius of the accretion disc,
$r_{out}$. There will be some trade-off between $r_{out}$ and cos(i)
which we did not explore in detail.

\begin{figure}
\begin{center}
\setlength{\unitlength}{1cm}
\begin{picture}(20,6)
\put(0.,-0.5){\includegraphics{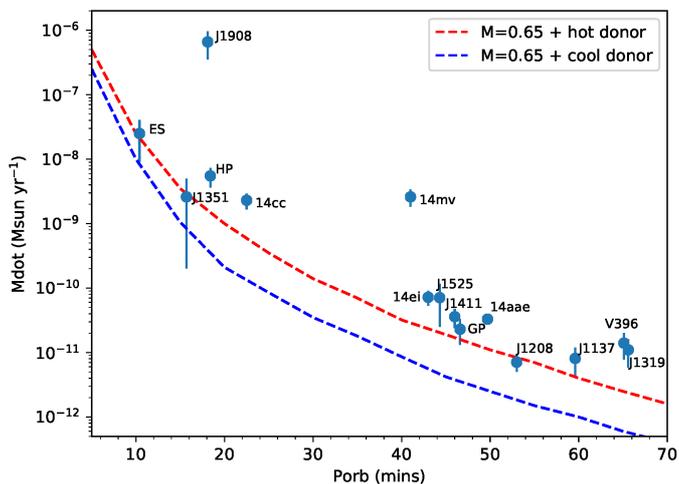}}
\end{picture}
\end{center}
\caption{The predicted mass accretion rate, $\dot{M}$, of
  \citet{Bildsten2006} as a function of orbital period, where the
  accretor is a 0.65\Msun white dwarf and the blue track has a cool
  donor star and the red track a hot donor star. We show the mass
  transfer rate derived using a white dwarf plus accretion disc model,
  GDR2 parallax data and multi-wavelength photometric measurements.}
\label{mdot-period-bildsten} 
\end{figure}

We use modules readily available in python modules including {\tt
  astropy}.  To determine the mass transfer rate and its uncertainty
we randomly select values for distance, mass and extinction within the
assigned errors with the other parameters kept fixed and repeat
  this 50 times. From these solutions we determine the mean value and
  standard uncertainty. Although this is a fairly simple approach the
mass transfer rate it is robust enough to make general conclusions
from the results as a whole. In Table \ref{mdot} we show the derived
mass transfer rates. In Figure \ref{mdot-period-bildsten} we show the
mass transfer rate as a function of orbital period along with two
predicted tracks from \citet{Bildsten2006} and we show the fit to the
SED's in Figure \ref{sed1} and Figure \ref{sed2}.

In Figure \ref{mdot-period-bildsten} we can see that some systems
appear to lie close to the predicted evolutionary track of a
$M_{1}$=0.65\Msun\hspace{1mm} and hot donor star. However, there is a
tendency for sources to lie above this track. In two cases, SDSS J1908
and ASASSN-14mv, their mass transfer rate is several orders of
magnitude greater than the hot donor track. SDSS J1908 lies above the
predicted absolute magnitude for its orbital period by $\sim$1 mag
(Figure \ref{PorbMv}), which is consistent with a higher mass transfer
rate. On the other hand, ASASSN-14mv has an absolute magnitude
consistent with expectations, which seems to be at odds with the high
mass transfer rate. For the eclipsing system Gaia14aae, we predict a
mass transfer rate of log10=--10.48, which is within 3$\sigma$ of the
prediction (log10=--10.74) based on using the absolute magnitude of
the bright spot (\S \ref{brightspot}. Despite these caveats, our
results show evidence that the mass transfer rate for the majority of
systems is higher than predicted from the models -- and in some cases
perhaps by an order of magnitude or more.

The most obvious reason for this result is that the donor star is
larger -- with a correspondingly higher mass transfer rate -- than
predicted. Indeed, recent observational work has shown that the donor
star is larger than expected from white dwarfs models and in some
cases larger than helium star models (\citep{Green2018b}). Moreover, a
detailed study of the eclipsing system Gaia14aae (\citep{Green2018a})
were able to determine system parameters ($M_{1}, M{_2}, R_{1},
R_{2}$) to high accuracy. This suggests that this system is likely to
have originated via the hydrogen CV evolutionary track. However, no
hydrogen is observed in its spectrum as would be expected. These
results taken with our findings should provide an impetus to
revisiting the formation models.

\begin{table*}
\begin{center}
\begin{tabular}{lrrrrr}
  \hline
Source & Period & log10($\dot{M}$  & $\pm$ & $\dot{M}$ & $\pm$ \\
& (mins) & ($M_{\odot}$ yr$^{-1}$) &   &  ($M_{\odot}$ yr$^{-1}$) \\
\hline
ES Cet      & 10.4 &  -7.61 &  0.46 &  $2.5\times10^{-8}$ & $1.6\times10^{-8}$   \\
SDSSJ1351   & 15.7 &  -8.58 &  1.13 &  $2.6\times10^{-9}$ & $2.4\times10^{-9}$   \\
SDSSJ1908   & 18.1 &  -6.18 &  0.27 &  $6.6\times10^{-7}$ & $3.1\times10^{-7}$   \\
HP Lib      & 18.4 &  -8.26 &  0.19 &  $5.5\times10^{-9}$ & $1.9\times10^{-9}$   \\
ASASSN-14cc & 22.5 &  -8.63 &  0.14 &  $2.3\times10^{-9}$ & $6.5\times10^{-10}$  \\
ASASSN-14mv & 41.0 &  -8.59 &  0.16 &  $2.6\times10^{-9}$ & $7.9\times10^{-10}$  \\
ASASSN-14ei & 43.0 & -10.14 &  0.13 &  $7.2\times10^{-11}$ & $1.9\times10^{-11}$ \\ 
SDSSJ1525   & 44.3 & -10.15 &  0.46 &  $7.1\times10^{-11}$ & $4.6\times10^{-11}$ \\
SDSSJ1411   & 46.0 & -10.44 &  0.17 &  $3.6\times10^{-11}$ & $1.2\times10^{-11}$ \\
GP Com      & 46.6 & -10.64 &  0.25 &  $2.3\times10^{-11}$ & $1.0\times10^{-11}$ \\
Gaia14aae   & 49.7 & -10.48 &  0.06 &  $3.3\times10^{-11}$ & $4.3\times10^{-12}$ \\
SDSSJ1208   & 53.0 & -11.15 &  0.15 &  $7.1\times10^{-12}$ & $2.1\times10^{-12}$ \\
SDSSJ1137   & 59.6 & -11.09 &  0.28 &  $8.1\times10^{-12}$ & $3.9\times10^{-12}$ \\ 
V396 Hya    & 65.1 & -10.86 &  0.26 &  $1.4\times10^{-11}$ & $6.2\times10^{-12}$ \\
SDSSJ1319   & 65.6 & -10.94 &  0.26 &  $1.1\times10^{-11}$ & $5.2\times10^{-12}$ \\
\hline
\end{tabular}
\caption{The mass transfer rates determined by fitting the SED of each
  system using a model consisting of a white dwarf plus steady-state
  accretion disc model. Apart from the eclipsing system (Gaia14aae), we
  fixed the mass of the white dwarf at $M_{1}=0.8\pm$0.1 \Msun and
  cos(i)=0.5. We varied radius of the accretion disc to achieve a good
  fit. The large error the mass transfer rate of SDSS J1505 is due to the large error on the distance.}
\label{mdot}
\end{center}
\end{table*}

\section{Space density of AM~CVn stars}

The AM~CVn stars are at the very shortest period end of the binary
star orbital period distribution and their space density is a
sensitive test for binary evolutionary models. Together with the
non-interacting double white dwarf binary, their number largely sets
the astrophysical background for gravitational wave detectors such as
{\sl LISA}. \citet{Nelemans2001a} found that the space density of
AM~CVn stars was very uncertain and subject to uncertainties due to
selection effects but determined a range of 0.4--1.7$\times10^{-4}$
pc$^{-3}$ (limits being termed `pessimistic' and `optimistic'
respectively).

With the advent of the large scale Sloan Digital Sky Survey (SDSS) it
was possible to obtain optical spectra of stars which had colours
consistent with the known AM~CVn stars and therefore identify new
systems in a systematic manner. Based on SDSS data
\citep{Roelofs2007b} found a space density of $1-3\times10^{-6}$
pc$^{-3}$ -- more than an order of magnitude lower than the
pessimistic model of Nelemans et al. (2001). Using a significantly
expanded SDSS sample, \citet{Carter2013a} derived an observed space
density of $5\pm3\times10^{-7}$ pc$^{-3}$, a value which is even lower
than that of \citet{Roelofs2007b}, and is currently the most reliable
estimate.

Perhaps the greatest source of uncertainty in predicting the space
density lies in the relative importance of the formation channels. As
we summarise in \S 1, there are three predicted channels -- detached
double white dwarfs that start mass transfer at very short orbital
periods; systems in which a non-degenerate core helium burning star
starts mass transfer and binaries in which a white dwarf is accreting
from a semi-degenerate star (the hydrogen CV channel). The very low
observed space density is compatible with only the double white dwarf
contribution to the formation. So either there are many fewer systems
coming from the helium star channel (and the CV channel) or we do not
sufficiently understand which detached binaries turn into stable
mass-transfer systems.

The main challenge in obtaining a space density estimate from the full
sample of known AM~CVn stars shown in Table~\ref{sources} is that it
contains systems discovered by a variety of different methods and
surveys. For example, variability, the presence of emission lines and
unusual colors have all been used to detect AM~CVn stars. As a result,
the sample in Table~\ref{sources} is neither flux- nor volume-limited,
but is instead affected by complex and often poorly understood
selection effects. Determining completeness corrections for the entire
sample with confidence is therefore impossible.

Nevertheless, the Gaia DR2 distances for the known AM~CVn stars do
contain crucial information. First, we can test if the sample is
complete out to some limiting distance. If so, then a space density
estimate can be derived immediately from this volume-limited
subsample. Second, even if completeness cannot be established out to
any distance, the sample can at least be used to set a firm lower
limit on the space density.

\begin{figure}
\begin{center}
\setlength{\unitlength}{1cm}
\begin{picture}(20,11.9)
\put(-0.2,-0.2){\includegraphics{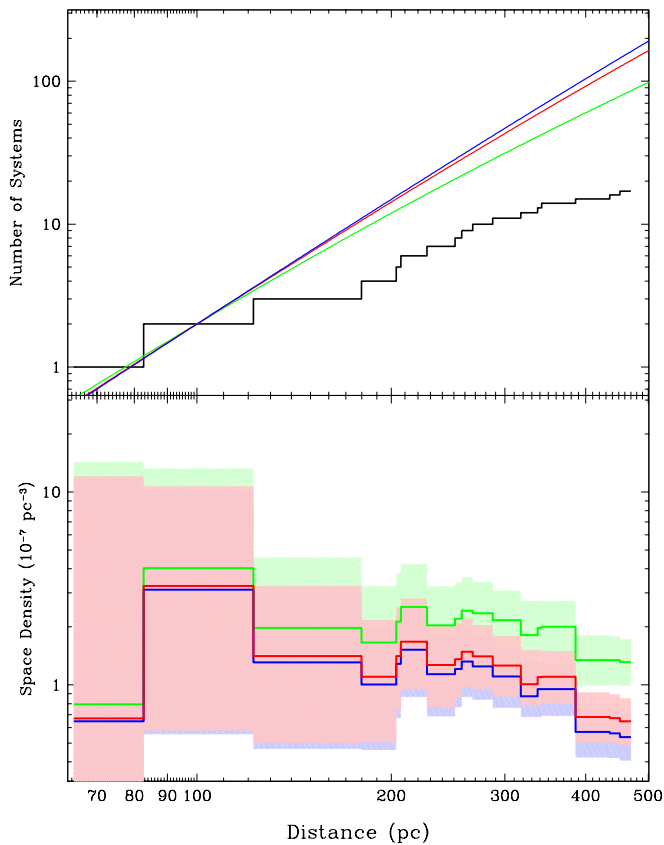}}
\end{picture}
\end{center}
\caption{{\em Top panel:} Observed and predicted cumulative
  distribution of AM~CVn stars as a function of distance. Predicted
  distributions are shown for Galactic disk models with three
  different exponential vertical scale heights: 100~pc (green), 300~pc
  (red) and 500~pc (blue). Given that AM~CVn stars are thought to be
  preferentially old systems, the two larger scale heights are likely
  to be more appropriate for this population. All models have been
  normalized to the data at 100~pc. The clear discrepancy between
  models and data -- which increases with distance -- points to
  serious incompleteness in the observational sample. {\em Bottom
    panel:} The space density (and 1$\sigma$ error bands) as a
  function of distance, as estimated from the number of AM~CVn stars
  and the effective volume contained within this distance. Effective
  volumes have been calculated for the same three Galactic disk models
  as shown in the top panel, thus providing three different space
  density estimates.}
\label{distance} 
\end{figure}

Figure~\ref{distance} shows the results of implementing these
ideas. The top panel shows the cumulative distribution of AM~CVn stars
as a function of distance, compared to the expected distributions for
stellar populations associated with the Galactic disk. It is
immediately obvious that completeness cannot be established with
confidence out to any distance. Based on this data set alone, it is
quite possible that there is a significant population of undiscovered
AM~CVn stars, with systems waiting to be discovered even within
$\simeq 100$~pc.

The bottom panel of Figure~\ref{distance} shows the space density one
would estimate from the number of AM~CVn stars as a function of
distance, $d$. This is given by $N/V_{eff}(d)$, where $N$ is the
number of systems found at distances smaller than $d$ and $V_{eff}$ is
the effective volume contained within $d$ in a given Galactic
model. Since our sample is likely to be incomplete everywhere, these
estimates must be treated as lower limits. The lower limit suggested
by our highest space density estimate would be $\rho_0 > 3 \times
10^{-7}$~pc$^{-3}$. However, allowing for the Poisson errors affecting
our small sample, our best 2$\sigma$ limit is substantially weaker,
$\rho_0 > 7 \times 10^{-8}$~pc$^{-3}$.

\section{Searching for companion stars to the AM~CVn stars}

We performed a search to identify companions to those AM~CVn stars
which have a {\sl Gaia} DR2 parallax determined to better than 20
percent. For each AM~CVn star, we obtained the {\sl Gaia} DR2 data for
all sources within 20$^{'}$, and then searched for stars which had a
parallax and proper motion in RA and DEC within 3$\sigma$.  Two AM~CVn
stars, KL~Dra and SDSS\,J1525+3600, had nearby stars which fulfilled
these criteria: their positions and properties are shown in Table
\ref{nearby}.

\begin{table*}
\caption{The results of searching for physically nearby stars to
  AM~CVn stars. The candidate companion stars were selected by
  searching for stars within 3$\sigma$ of the parallax to the AM~CVn
  star and also 3$\sigma$ of the proper motions in RA and DEC.  In the
  final column we give the physical separation between the star and
  the AM~CVn star.}  \resizebox{\textwidth}{!}{ \centering
  \begin{tabular}{lcrrrrrrrr}
    \hline
     & {\sl Gaia}  & RA    & Dec   & p     & pmra     & pmdec    & G     & g-r & Separation\\
     & SourceID    & J2000 & J2000 & (mas) & (mas/yr) & (mas/yr) & (mag) &     & (pc) \\
     \hline
KL\,Dra & 2239471475135041664 & 291.159333 & 59.696233 & 1.04(15) & -2.46(30) & -18.27(26) &  19.05 &   1.21 & \\
       & 2239448930851818112 & 290.953649 & 59.610338 & 1.11(33) & -2.89(75) & -18.37(80) &  19.44 &   1.14 & 3.5$\pm$0.6 \medskip\\
SDSS\,J1525+3600  & 1375131155313563136 & 231.289887 & 36.015117 & 1.93(28) & 4.79(81)  & -16.87(81)   &  19.665 &  0.60 & \\
                & 1375135415921155968 & 231.150531 & 36.107433 & 1.63(60) & 5.25(129) & -16.410(129) &  20.226 &  1.13 & 1.4$\pm$0.3 \\
		& 1375131155313567232 & 231.288185 & 36.023225 & 2.28(15) & 5.24(28)  & -14.492(28)  &  18.293 &  1.20 & 0.06$\pm$0.01 \\
\hline
\label{nearby}
\end{tabular}}
\end{table*}

We determined the physical separation between the candidate companion
stars and show these in the final column of Table \ref{nearby}. The
candidate companion star to KL~Dra is 12.4$^{'}$ distant (=3.5 pc),
whilst for SDSSJ1525+3600 they are 8.4$^{'}$ and 6.1$^{''}$ (=1.4 pc
and 0.06 pc or 1.25$\times10^{5}$ AU). The more distant candidate
companions are too distant to have any effect on the dynamics of the
AM~CVn binary, although they may have been much closer in the distant
past. However, the candidate companion to SDSSJ1525+3600 is only
6.1$^{''}$ distant providing powerful impetus to obtain a radial
velocity measurement to determine whether it is physically associated
with the AM~CVn binary. If they do show similar velocities then they
will be interesting from a binary formation and evolutionary point of
view.

\section{Conclusions}

For the first time we have reliable distances to more than a few
AM~CVn stars. Using these distances we determine the expected
cumulative distribution of AM~CVn stars and compare it with the
distribution of a Galactic disk population. We find that there is
likely to be a significant number of AM~CVn stars awaiting to be
discovered. Since we find that the location of AM~CVn stars are in a
distinctive region of the HR diagram (Figure \ref{hrdiag}) the {\sl
  Gaia} DR2 dataset will be a useful tool for identifying candidate
systems based on their colour and absolute magnitude.

One of the great uncertainties in predicting the space density of
AM~CVn stars has been in determining the relative importance of the
three formation channels. There is now mounting evidence that the
majority donor stars in AM CVn stars are not fully
  degenerate. We were able to determine the mass transfer rate in 15
AM CVn stars and find that most have rates which are greater than
predicted by standard tracks. The donors appear to be larger than
expected.

We also find that none of the AM~CVn stars which have proper motion
and parallax data are likely to be halo objects. Those objects which
have very low heavy element abundances are therefore not likely to be
due to their age. Coupled with the findings that the parameters of the
eclipsing AM~CVn star, Gaia14aae, does not readily fit with the models
should serve as impetus to revisit these models and the nature of the
donor star in particular.

\section{Acknowledgements}

This work has made use of data from the European Space Agency (ESA)
mission {\it Gaia} (\url{https://www.cosmos.esa.int/gaia}), processed
by the {\it Gaia} Data Processing and Analysis Consortium (DPAC,
\url{https://www.cosmos.esa.int/web/gaia/dpac/consortium}). Funding
for the DPAC has been provided by national institutions, in particular
the institutions participating in the {\it Gaia} Multilateral
Agreement. We extracted Galex data from Multi-Mission archive at the
Space Telescope Science Institute (MAST). The Pan-STARRS1 Surveys
(PS1) and the PS1 public science archive have been made possible
through contributions by the Institute for Astronomy, the University
of Hawaii, the Pan-STARRS Project Office, the Max-Planck Society and
its participating institutes, the Max Planck Institute for Astronomy,
Heidelberg and the Max Planck Institute for Extraterrestrial Physics,
Garching, The Johns Hopkins University, Durham University, the
University of Edinburgh, the Queen's University Belfast, the
Harvard-Smithsonian Center for Astrophysics, the Las Cumbres
Observatory Global Telescope Network Incorporated, the National
Central University of Taiwan, the Space Telescope Science Institute,
the National Aeronautics and Space Administration under Grant
No. NNX08AR22G issued through the Planetary Science Division of the
NASA Science Mission Directorate, the National Science Foundation
Grant No. AST-1238877, the University of Maryland, Eotvos Lorand
University (ELTE), the Los Alamos National Laboratory, and the Gordon
and Betty Moore Foundation. The national facility capability for
SkyMapper has been funded through ARC LIEF grant LE130100104 from the
Australian Research Council, awarded to the University of Sydney, the
Australian National University, Swinburne University of Technology,
the University of Queensland, the University of Western Australia, the
University of Melbourne, Curtin University of Technology, Monash
University and the Australian Astronomical Observatory. SkyMapper is
owned and operated by The Australian National University's Research
School of Astronomy and Astrophysics. The survey data were processed
and provided by the SkyMapper Team at ANU. The SkyMapper node of the
All-Sky Virtual Observatory (ASVO) is hosted at the National
Computational Infrastructure (NCI). Development and support the
SkyMapper node of the ASVO has been funded in part by Astronomy
Australia Limited (AAL) and the Australian Government through the
Commonwealth's Education Investment Fund (EIF) and National
Collaborative Research Infrastructure Strategy (NCRIS), particularly
the National eResearch Collaboration Tools and Resources (NeCTAR) and
the Australian National Data Service Projects (ANDS).  This research
made use of Astropy, a community-developed core Python package for
Astronomy \citep{astropy2018}. Armagh Observatory and Planetarium is
core funded by the Northern Ireland Executive through the Dept for
Communities.  MJG acknowledges funding from an STFC studentship via
grant number ST/N504506/1.  A.A. acknowledges the support of the
National Research Council of Thailand (grant number R2561B087).

%----------------------------------------------------------------------
% BIBLIOGRAPHY - Please add additional references to the file biblio.tex
%----------------------------------------------------------------------

%----------------------------------------------------------------------
\appendix
%----------------------------------------------------------------------

%----------------------------------------------------------------------
% APPENDIX A
%----------------------------------------------------------------------
\section{Photometry of AM~CVn stars: online table}

We provide an online table in fits
format\footnote{\url{https://www.ast.cam.ac.uk/\textasciitilde{}ebreedt/web/AMCVn_datatable_Ramsay_etal_2018.fits}}
  which includes data on all of the known AM~CVn stars.  The columns
  and caveats are described below.

\subsection{Target names and coordinates}
In the main body of the paper (e.g. Table~\ref{sources}), we use IAU
Variable Star names or shortened survey IDs for each of the
targets. The online table includes these IDs (col.\,1), as well as
alternative names by which each target is known (col.\,2).

We give the J2015.5 coordinates from \Gaia\, where available
(cols.\,5,~6), as well as ICRS J2000.0 coordinates for all sources
(cols.\,3,~4). The J2000.0 coordinates were derived from the Gaia
5-parameter astrometry where available, otherwise the table shows the
most precise coordinates available for that target. We also computed
the IAU 1958 Galactic latitude and longitude for each target. These
are given in columns\,7 and 8.

The \Gaia\, DR2 source ID is shown in col.\,9.

\subsection{Orbital period and brightness range}

Cols.\,10 and 11 show the period of the binary in days and minutes
respectively, with a flag/comment in col.\,12 to indicate whether the
period is the orbital period (`orb', most reliably measured from time
series spectroscopy), the superhump period (`sh', resulting from disc
precession during the outburst and typically a few per cent longer
than the orbital period), estimated from the outburst recurrence time
(`est rec', \citealt{Levitan2015}) or whether no period has been
measured for the system so far (`unknown'). References for the
discovery and the period measurements are indexed in col.\,15,
corresponding to the following references:

\begin{small}
\noindent 
 [1] \citealt{RCH2002},
 [2] \citealt{Ramsay2002},
 [3] \citealt{Espaillat2005},
 [4] \citealt{Nelemans2001b}, 
 [5] \citealt{Fontaine2011},
 [6] \citealt{Kupfer2015},
 [7] \citealt{ODonoghue1994},
 [8] \citealt{Levitan2014},
 [9] \citealt{Wevers2016},
[10] \citealt{Kato2015},
[11] \citealt{Wood1987},
[12] \citealt{Wood2002},
[13] \citealt{Kato2004},
[14] \citealt{Levitan2015},
[15] \citealt{Levitan2011},
[16] ASASSN \url{http://www.astronomy.ohio-state.edu/~assassin/transients.html},
[17] AAVSO \url{https://www.aavso.org},
[18] \citealt{Copperwheat2011},
[19] \citealt{Abbott1992},
[20] \citealt{Levitan2013},
[21] \citealt{Thorstensen2012},
[22] \citealt{Motsoaledi2015},
[23] \citealt{WoodVasey2003},
[24a] \citealt{Carter2014a},
[24b] \citealt{Carter2014b},
[25] \citealt{Woudt2005},
[26] \citealt{Shears2012},
[27] \citealt{Roelofs2005},
[28] \citealt{Roelofs2009},
[29] \citealt{Anderson2005},
[30] \citealt{Ramsay2012},
[31] \citealt{Rau2010},
[32] \citealt{Nather1981},
[33] \citealt{Prieto2014},
[34] \citealt{Woudt2013},
[35] \citealt{Kato2014},
[36] \citealt{Campbell2015},
[37] \citealt{Marsh2017},
[38] CRTS \url{http://nesssi.cacr.caltech.edu/DataRelease},
[39] \citealt{Anderson2008},
[40] \citealt{Kupfer2013},
[41] \citealt{Kepler2015},
[42] \citealt{Wagner2014},
[43] \citealt{Roelofs2007a},
[44] \citealt{Ruiz2001},
[45] \citealt{Green2018a},
[46] Kato, vsnet-alert 18124,
[47] Aungwerojwit et al. (in prep),
[48] Breedt et al. (in prep),
[49] Kupfer et al. (in prep),
[50] Maehara \& Kojima, vsnet-alert 22174,
[51] \citealt{Green2018_14aae},
[52] \citealt{Kong2018}
\label{table1ref}
\end{small}

The Y/N flag in col.\,13 indicates whether the system has been observed in outburst, and the magnitude range (with filter information) is given in col.\,14.

\subsection{Parallax, reddening and derived quantities}

The calculation of the following quantities were discussed in detail in the main text:\\ 
Col.\,16-17: Parallax and error (mas), taken from \Gaia\, DR2\\
Col.\,18-19: Distance and error (pc) --- calculated from an exponentially decreasing space density prior with $L=400$~pc. See Section~\ref{sec:distances}.\\
Col.\,20-21: Extinction and reddening parameters $E(B-V)$ and $A_g$ towards each system (mag), derived from the Pan-STARRS1 distance reddening maps \citep{Green2018b} as discussed in Section~\ref{sec:ext}. ASASSN-14cc, ASASSN-14ei, NSV1440 and Gaia16all have parallax measurements from Gaia, but are not in the Pan-STARRS1 footprint. For these four stars we show the maximum reddening from \citet{Schlafly2011}. A note has been added in col.\,66 of the table.\\
Col.\,22: $g$ band quiescent magnitude taken from Pan-STARRS, used to calculated the absolute magnitude\\
Col.\,23-24: absolute magnitude $M_g$ and its associated error (mag)

\subsection{\galex}
\label{Galex}
Although \Swift\, and \xmm\, has been used to observe some AM~CVn
stars, \galex\, provided the opportunity to take an unbiased snapshot
survey of their near UV flux.  \galex\, was launched in April 2003 and
had a lifetime of ten years and performed observations at UV
wavelengths \citep{Martin2005}. It had a primary mirror 50~cm in
diameter and had a field of view of 1.25~degrees. It had two UV
channels, the Near UV (NUV, central wavelength 2271\AA) and the Far UV
(FUV, 1528\AA), the latter failing after 6 years. Whilst it could
perform relatively short pointed observations, it also carried out a
wide field survey and by its end it had sampled 3/4 of the sky in one
UV band \citep{Bianchi2014}. 41 of the known AM~CVn stars are included
in the catalogue of \citet{Bianchi2014}. We show the FUV and NUV
fluxes in units of mJy in cols.\,25--26 (FUV) and cols.\,27--28 (NUV)
of the online table.

\subsection{SDSS}
\label{SDSS}

The Sloan Digital Sky Survey (SDSS) uses a 2.5m telescope at Apache
Point Observatory in New Mexico, and has a field of view 3 degrees
across \citep{York2000}. Its Data Release 7 includes a photometric
catalog of northern-hemisphere targets \citep{Abazajian2009}.

We performed a cross-match of known AM~CVn J2000.0 positions with all
SDSS photometric objects, using a search radius of 2 arcsec. We
identified 34 AM~CVns which have SDSS photometry. All but two of these
magnitudes are consistent with the faint end of the range shown in
Table~\ref{sources}. The two outliers are SDSS\,J0804+1616, which is
brighter by approximately 0.7~mag \citep[perhaps consistent with the
  large-scale variability this system shows even in
  quiescence;][]{Roelofs2009} and PU\,Aqr=SDSS\,J2047+0008, which is
brighter by approximately 1.8~mag and may have been in outburst or in
decline from outburst when the SDSS data were collected.

The SDSS $ugriz$ magnitudes and their errors are shown in columns
29--38 of the online table.

\subsection{Pan-STARRS1}
\label{Panstarrs}

Pan-STARRS1 is a 1.8m telescope with a field of view of 7 square
degrees, located on Halakala Observatory in Hawaii. There are five
broadband filters which are close to, but not exactly, the SDSS
filters $grizy$ \citep{Chambers2016,Tonry2012}. It is performing an
all-sky survey with declinations $>-30^{\circ}$.

We performed a cross-match with Pan-STARRS1 DR1 using a 2~arcsec match
radius to account for uncertainty in the recorded positions of the
AM~CVn stars. The mean AB PSF $grizy$ magnitudes are shown in
columns~39--48 of the online table for all systems which were found to
have a match.

For those AM~CVn stars which do not show outbursts the mean $g$ mag is
consistent with the values shown in Table \ref{sources}. For those
systems which do show outburst, Pan-STARRS1 has a mean magnitude which
is consistent with the source being in quiescent apart from three
systems: CR\,Boo where $g$ is more consistent with it being in
quiescent and the other filters in outburst; SDSS\,J0804+1616 has
range in magnitude in Table \ref{sources} between 17.8--19.0 and the
mean Pan-STARRS1 mag appears in the middle of this range and
CRTS\,J0744+3254 which has a Pan-STARRS1 $g$ mag much fainter than
$r$. The Pan-STARRS1 DR1 gives a mean magnitude of the source over the
pointings covered. However, it also gives a range in magnitudes for
the different filters. For the three sources which show evidence for
being in different accretion states, we show in
Table~\ref{panstarrsextra} the range in these filters' measurements.

\begin{table*}
\caption{The PanSTARRS1 measurements of these three AM~CVn stars
  appears to have been taken in different accretion states in the
  different bands. In addition to the magnitudes shown in the online
  table (and repeated here for clarity), we also show the range in
  magnitude for the detections in each filter.\label{panstarrsextra}}
  \begin{tabular}{lccccc}
    \hline
Source & $g$ &  $r$ &  $i$ &  $z$ &  $y$ \\
\hline
  CR\,Boo                   &  16.630(42) & 14.406(130)& 14.965(92) & 15.154(142)& 15.494(96) \\
  {\sl \hspace{5mm}range} &  16.61..16.78 &  14.12..16.74 &  14.56..15.65 &   14.47..17.07 &  15.27..17.41 \smallskip\\
  SDSS\,J0804+1616          &  18.222(61) & 18.432(44) & 18.556(26) & 18.639(33) & 18.493(29) \\
  {\sl \hspace{5mm}range} &  18.09..18.60 &  18.21..18.60 &  18.50..18.77 &   18.50..18.86 &  18.25..18.92 \smallskip\\           
  CRTS\,J0744+3254          &  21.345(275)& 18.884(21) & 21.397(55) & 17.803(23) & 18.298(16) \\
  {\sl \hspace{5mm}range} &  21.32..21.87 &  18.81..18.91 &  21.30..21.50 &   17.81..17.89 &  18.29..18.32 \\           
  \hline
  \end{tabular}
\end{table*}

\subsection{Skymapper}
\label{Skymapper}

Skymapper is a telescope with a 1.35m primary mirror located Siding
Spring Observatory, Australia, and has a camera which covers
2.4$\times$2.3 degrees of the sky \citep{Scalzo2017}. Its primary goal
is to obtain a map of the southern sky in a number of filters, with
periods of poor conditions being focussed on detecting
transients. There are six filters, with the bluest $u,v$, being narrow
band filters (FWHM=42 and 28~nm respectively) which cover the SDSS~$u$
band. The filters $griz$ differ compared to their SDSS equivalents by
40~nm in effective wavelength and width \citep{Bessell2011}.

We performed a 2~arcsec radius cross-match with the Skymapper DR1.1
release and find that six AM~CVn stars have at least one detection in
one filter. We show the magnitudes in the $uvgrzi$ filters on the AB
mag scale in columns~49--60 of the online table.

\subsection{Kinematics}
\label{kinematics}

Columns 61--64 list the proper motions in RA and Dec (mas) from
\Gaia\, DR2. The population membership (thick/thin disc of the
Galaxy), as computed in Section~\ref{sec:galdist}, is shown in
col.\,65. For convenience, these data are also shown in
Table~\ref{population}.

The final column in the online table, 66, is a text field for specific notes about the data for that target.

% APPENDIX B

\newpage
\section{Spectral Energy Distributions and Kinematic data}

\begin{figure*}
\caption{The spectral energy distribution of AM~CVn stars which
    have a parallax accurate to within 20 percent and have
    multi-colour survey data in more than two filter passbands. In the
    solid line we show the model spectral energy distribution where
    $M_{1}$, $E_{B-V}$ and distance are restricted to the values shown
    in each panel and the mass transfer rate, $\dot{M}$, is a free
    parameter.}
\begin{center}
\setlength{\unitlength}{1cm}
\begin{picture}(16,20)
\put(1,0){\includegraphics{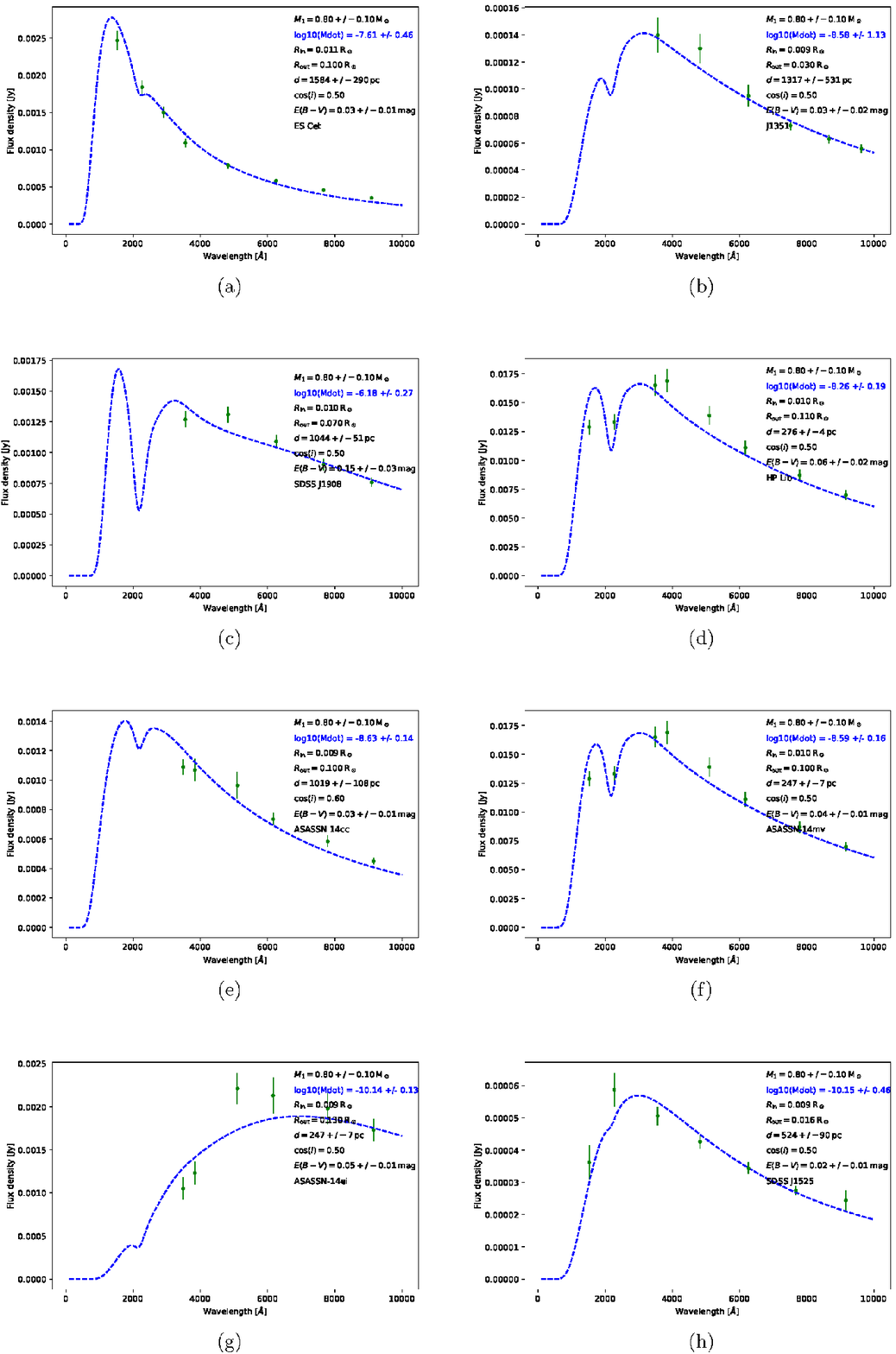}}
\end{picture}
\end{center}
\label{sed1}
\end{figure*}

\begin{figure*}
\caption{Continued from Figure B1.} 
\begin{center}
\setlength{\unitlength}{1cm}
\begin{picture}(16,20)
\put(1,0){\includegraphics{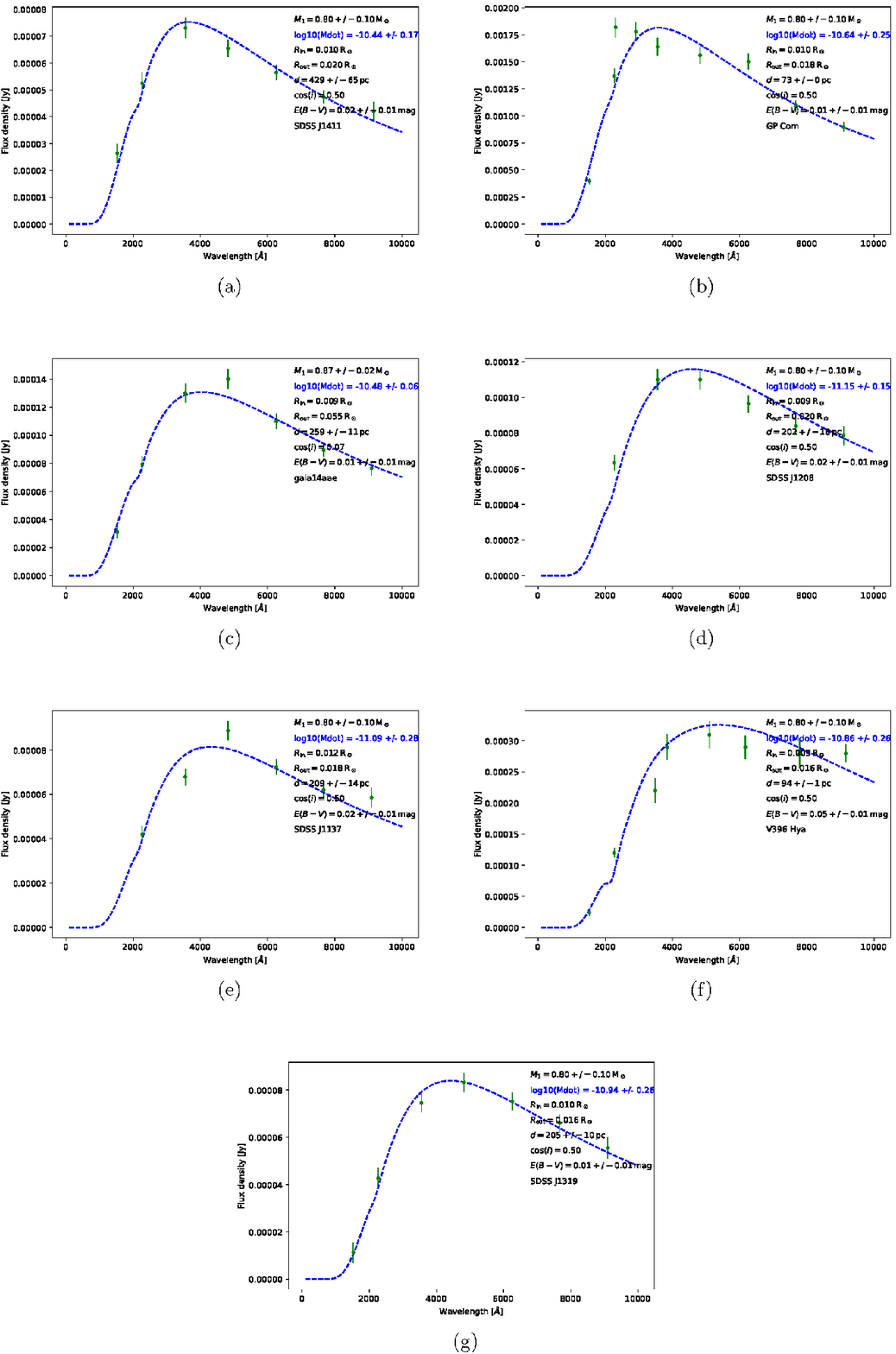}}
\end{picture}
\end{center}
\label{sed2}
\end{figure*}

\begin{table*}
\caption{Kinematic information for the AM~CVn stars with measured
  proper motions in \Gaia\, DR2. These data are also included in the
  online table. \label{population}}
\begin{center}
  \begin{tabular}{lrrl}
    \hline
    Source   &  pmra     & pmdec & population  \\
             & (mas/yr) & (mas/yr)  &  \\
\hline
V407 Vul          &  $-2.315\pm0.386$ &  $-3.726\pm0.719$ &  thin \\
ES Cet            &  $17.486\pm0.196$ &  $-3.075\pm0.157$ & thick \\
SDSS J1351-0643   &   $5.503\pm0.586$ & $-17.328\pm0.465$ &   thick \\
AM~CVn           &  $30.935\pm0.073$ &  $12.420\pm0.053$ &  thick \\
SDSS J1908+3940   &  $-7.019\pm0.087$ &  $-5.332\pm0.093$ & thin \\
HP Lib            & $-28.776\pm0.099$ & $-12.636\pm0.074$ &  thin \\
PTF1 J1919+4815   &   $5.548\pm0.535$ &   $5.850\pm0.830$ & thin/thick \\
CX361             &   $2.393\pm0.344$ &  $-2.001\pm0.288$ & thin \\
ASASSN-14cc       & $-14.897\pm0.115$ & $-37.800\pm0.111$ & thick \\
KL~Dra            &  $-2.463\pm0.296$ & $-18.269\pm0.262$ & thick \\
PTF1 J0719+4858   &   $4.276\pm0.297$ &   $2.345\pm0.166$ & thin \\
YZ LMi            & $-30.506\pm0.841$ &  $-4.072\pm1.084$ & thin \\
CP Eri            &  $18.090\pm1.445$ &  $-7.174\pm1.550$ & thin \\
SDSS J1043+5632   &  $-0.561\pm0.936$ &  $-2.997\pm1.099$ & thin  \\
CRTS J0910-2008   &   $3.730\pm1.050$ &  $-8.610\pm0.923$ & thin/thick \\
CRTS J0105+1903   &   $7.159\pm1.023$ &   $8.640\pm0.923$ & thin \\
V406 Hya          &  $-1.429\pm1.606$ &  $-6.191\pm1.389$ & thin \\
SDSS J1730+5545   &   $4.104\pm0.829$ & $-33.861\pm0.931$ & thick \\
NSV1440           &   $7.737\pm0.255$ &  $27.446\pm0.277$ & thin \\
V558 Vir          & $-13.989\pm2.482$ &  $-3.959\pm3.140$ & thin \\
SDSS J1240-0159   &  $-1.852\pm1.248$ &  $-8.006\pm0.646$ & thin \\
V744 And          &  $-2.691\pm0.827$ &   $6.332\pm1.797$ & thick  \\
SDSS J1721+2733   &   $2.815\pm0.923$ &  $-2.025\pm1.102$ & thin \\
ASASSN-14mv       & $-17.270\pm0.266$ & $-43.836\pm0.205$ & thin \\
ASASSN-14ei       & $-10.263\pm0.074$ & $-15.380\pm0.098$ & thin  \\
SDSS J1525+3600   &   $4.787\pm0.518$ & $-16.876\pm0.810$ & thin \\
SDSS J0804+1616   &   $1.365\pm0.382$ &  $-5.972\pm0.219$ & thin \\
SDSS J1411+4812   & $-17.297\pm0.431$ &  $36.323\pm0.444$ & thick \\
GP~Com           & $-344.791\pm0.131$ &  $34.743\pm0.082$ & thin/thick \\
SDSS J0902+3819   & $-15.799\pm1.170$ & $-13.360\pm1.058$ & thin  \\
Gaia14aae         &  $-3.927\pm0.317$ & $-14.099\pm0.269$ & thin \\
SDSS J1208+3550  & $-112.865\pm0.651$ & $-56.236\pm0.510$ & thick \\
SDSS J1642+1934  &   $-2.560\pm1.034$ & $-28.606\pm1.082$ & thick \\
SDSS J1552+3201  &    $5.943\pm0.947$ & $-27.253\pm1.174$ & thin \\
SDSS J1137+4054  &   $23.118\pm0.383$ & $-56.566\pm0.336$ & thin \\
V396 Hya         & $-286.331\pm0.306$ & $-59.209\pm0.213$ & thick  \\
SDSS J1319+5915  &  $-27.885\pm0.498$ &  $12.375\pm0.359$ & thin \\
CRTS\,J0844-0128 &   $-5.291\pm1.621$ &  $-4.794\pm0.549$ & thin \\
SDSS J1505+0659  &   $44.487\pm0.724$ & $-25.417\pm0.686$ & thin/thick \\
Gaia\,16all      &   $-2.640\pm1.647$ &   $5.287\pm1.465$ & thin \\
\hline
\end{tabular}
  \end{center}
\end{table*}

\end{document}